\documentclass[fleqn,usenatbib]{mnras}

\usepackage{graphicx}	
\usepackage{amssymb}	
\usepackage{amsmath} 
\usepackage{hyperref}
\usepackage[hyphenbreaks]{breakurl}
\usepackage{diagbox, multirow, booktabs}
\usepackage{epsfig}

\newcommand{\secref}[1]{Section \ref{#1}}
\newcommand{\appref}[1]{Appendix \ref{#1}} 

\newcommand{\figref}[1]{Figure \ref{#1}}
\renewcommand{\eqref}[1]{Eq. (\ref{#1})}

\newcommand{\sex}{SExtractor}

\newcommand{\mstar}{$M_\ast$}

\newcommand{\lgmstar}{$\log_{10}(M_\ast/h^{-2}M_\odot)$}
\newcommand{\sersic}{S\'ersic}

\DeclareMathOperator{\sech}{sech}

\title[Superthin galaxies from optical to near-infrared]{Optically Selected Superthin Galaxies Remain Thin in the Near-infrared}

\author[J. Hu, C. Li \& D. D. Xu]{Jianhong Hu,$^{1,2}$\thanks{E-mail: jhu24@zzu.edu.cn}, Cheng Li,$^{2}$\thanks{E-mail: cli2015@tsinghua.edu.cn} and Dandan Xu$^{2}$\\
$^{1}$School of Physics and Laboratory of Zhongyuan Light, Zhengzhou University, Zhengzhou 450001, China\\
$^{2}$Department of Astronomy, Tsinghua University, Beijing 100084, China}

\date{Accepted XXX. Received YYY; in original form ZZZ}

\pubyear{2026}

\begin{document}
\label{firstpage}
\pagerange{\pageref{firstpage}--\pageref{lastpage}}
\maketitle

\begin{abstract}
We investigate whether galaxies identified as superthin in optical images remain superthin in the near-infrared (NIR), and how their extreme disk morphology is related to environment. From a nearby volume-limited sample, we select 210 superthin galaxies using two-dimensional bulge/disk decomposition of SDSS $r$-band images, requiring the disk component to have a minor-to-major axis ratio $b/a<1/9$. We measure disk shapes from SDSS $griz$ to UKIDSS $JHK$ bands. Both the major- and minor-axis scales decrease from the optical to the NIR, reaching $\sim0.6$ of their $r$-band values in the $K$ band, but the disk axis ratio remains nearly unchanged. Thus, optically selected superthin galaxies remain superthin in the NIR, suggesting that any NIR-dominant thick component, if present, is not prominent enough to alter the fitted disk shape. Reanalysis of our sample and a previous superthin sample shows that earlier reported NIR thickening is mainly due to a magnitude- and band-dependent bias in one-dimensional fitting. We further compare their environments with matched control samples using projected cross-correlations, reconstructed local overdensities, and large-scale-structure classifications. Superthin galaxies show lower clustering on $\sim0.1$--$1\,h^{-1}\,\mathrm{Mpc}$ scales and lower overdensities at $1\,h^{-1}\,\mathrm{Mpc}$, but similar large-scale clustering and no clear residual dependence on large-scale-structure type. These results are consistent with a higher central-galaxy fraction and lower satellite fraction in the superthin sample, while the similar large-scale clustering points to comparable typical halo masses relative to the controls. This halo-occupation interpretation is qualitatively consistent with simulation-based scenarios in which high host-halo spin helps build extended thin disks, although the present data do not directly measure halo spin.
\end{abstract}




\section{Introduction}
\label{sec:intr}

Superthin galaxies were first identified approximately six decades ago as a distinct class of edge-on spiral galaxies exhibiting exceptionally small disk axial ratios, with minor-to-major axis ratios typically exceeding $1:9$ and sometimes reaching as low as $1:30$ \citep{voron1967, devau1974, goad1979, goad1981}. These galaxies are characterized by the absence of a prominent bulge component and display strikingly flat stellar disks with minimal vertical thickness \citep{matthews1999}. Early spectroscopic studies revealed that superthins possess low emission-line ratios indicative of low metallicity and slow star formation, alongside slowly rising rotation curves that imply modest central mass concentrations \citep{goad1981}.

The intrinsic shapes of galaxies can be inferred statistically from their projected images by assuming a single ellipsoid and random orientation of the line of sight. Such studies indicate that the thickness-to-diameter ratio of typical galaxies follows a distribution with a mean of $0.21$ and a scatter of $0.05$ \citep[e.g.,][]{holmberg1946, sand1970, noerdlinger1979, binney1981, padilla2008}, corresponding to a minor-to-major axis ratio of approximately $1:5$ when viewed edge-on. When the disk component is isolated through two-dimensional decomposition, the mean minor-to-major axis ratio reduces to $\sim 1/8.9$ \citep{guthrie1992, degrijs1998, kregel2002, mosenkov2015}, yet this remains thicker than that of the majority of superthin galaxies. Superthins thus occupy the extreme flat tail of the axis ratio distribution, reflecting their classification as very late-type spirals \citep{heidmann1972}.

Disk galaxies typically comprise two distinct stellar components: a thin disk and a surrounding thick disk \citep[e.g.,][]{yoachim2006, comeron2011, martinez2019}. The thick disk is characterized by a scale height approximately four times larger than that of the thin disk, along with older, more metal-poor stellar populations. Superthin galaxies appear to represent an extreme case with very weak bulges and little evidence for a thick component that dominates their global disk shape, effectively constituting a nearly ``naked'' thin disk. This unique morphology naturally invites comparison with low surface brightness galaxies (LSBs), with which superthins share many physical properties, differing primarily in inclination: superthins are viewed edge-on while LSBs are observed face-on \citep{banerjee2017, narayanan2022}. Both classes exhibit high gas fractions, low star formation rates, and are dominated by dark matter in their outer regions \citep{matthews1999, bizyaev2021}.

The formation and persistence of superthin morphologies should in principle provide interesting constraints and challenges for galaxy formation models. In the current theoretical framework, galaxies form from condensed gas in the potential wells of dark matter halos \citep{white1978,fall1980,blumenthal1984,mo1998}. Angular momentum, acquired through tidal torques \citep{peebles1969,doroshkevich1970,white1984}, plays a crucial role in determining disk sizes \citep{fall1980,mo1998}. More generally, the retained fractions of specific angular momentum in stars, atomic gas, and baryons are closely linked to galaxy morphology and disk stability \citep{romeo2023}. Energy feedback prevents overcooling and helps preserve angular momentum in the stellar component \citep{weil1998,sommerlarsen1999,thacker2000}. Subsequent hierarchical clustering reshapes galaxies through mergers. While major mergers can drive morphological transformations \citep{toomre1972,toomre1977}, even minor mergers can dynamically heat disks or drive gas inflows, leading to thickening \citep{quinn1986,quinn1993,toth1992,walker1996,velazquez1999}. In extreme cases, successive minor mergers may even convert disk galaxies into ellipticals \citep{bournaud2007}. The observed abundance of thin disks has therefore been argued to pose a population-level challenge to cosmological simulations, which can produce too few intrinsically thin stellar disks \citep{haslbauer2022}. However, recent simulations reveal that the outcome of individual mergers depends sensitively on gas fraction, mass ratio, and orbital geometry; disk-dominant galaxies can survive even major mergers when gas-rich and on prograde coplanar orbits \citep{springel2005,athanassoula2016}. By tracking the evolution of superthin galaxies in the TNG100 simulation \citep{marinacci2018, naiman2018, nelson2018, pillepich2018, springel2018}, \citet{hu2024} found that the progenitors of present-day superthin galaxies had similar morphologies to normal disk galaxies at higher redshifts, developing more extended, flatter structures (smaller $b/a$) since $z \sim 1$ through frequent prograde mergers that significantly increase the spin of their dark matter halos.

To unravel the nature of superthin galaxies, numerous observations have been carried out. Individual objects such as UGC~7321 have been examined in detail \citep{goad1981,matthews1999}, revealing moderate ionization and low metallicity. \citet{karachentsev1989} found a tight Tully-Fisher relation for superthin galaxies, suggesting they may serve as distance indicators. More recently, large samples have enabled statistical investigations. Early efforts relied on visual selection \citep{karachentsev1993,karachentsev1999}. Kautsch and collaborators \citep{kautsch2006,kautsch2009a,kautsch2009b,kautsch2009c,kautsch2009d} constructed a sample of 15,127 edge-on galaxies from the Sloan Digital Sky Survey \citep[SDSS;][]{york2000} using an automated method, identifying a bulge-less (simple) disk fraction of 15\% and finding such systems in environments ranging from isolated fields to moderate-density groups. Bizyaev et al. \citep{bizyaev2014} built another edge-on galaxy sample and applied a one-dimensional photometric decomposition technique \citep{bizyaev2002,bizyaev2009} to measure disk radial-to-vertical scale ratios. From this sample, they selected 85 superthin objects with
disk radial-to-vertical scale ratios exceeding 9
\citep[][hereafter B17]{bizyaev2017} and found that they preferentially inhabit low-density environments with weaker connections to filaments. Subsequently, 49 of these galaxies were observed in the near-infrared \citep[NIR;][hereafter B20]{bizyaev2020}. Comparing optical and NIR structural parameters, B20 found that the NIR scale length is significantly shorter than the optical scale length, while the scale height remains relatively comparable between bands. Follow-up spectroscopy by \citet{bizyaev2021} revealed that a large fraction of blue superthin galaxies are dynamically underevolved, with low vertical velocity dispersions in both gas and stars. Recent theoretical work further suggests that low-surface-brightness ultrathin disks can remain thin in isolation because their low stellar surface density suppresses bar-driven heating, whereas models with larger stellar mass fractions can form bars and thicken substantially \citep{aditya2025}.

The result from B20 that superthin galaxies appear thicker in the NIR than in the optical raises a fundamental question: are superthin galaxies identified in the optical still superthin when observed in the NIR? This question directly addresses the structure of the stellar disk. If NIR images show a substantially thicker fitted disk, it would suggest that the older stellar populations contributing strongly to the NIR light are distributed over a larger vertical scale than the younger populations traced more strongly by optical light. Conversely, if the fitted disk shape remains nearly invariant from the optical to the NIR, the NIR light does not reveal a dominant thick component. Such a result would still allow more than one physical interpretation, including a relatively quiescent heating history, low internal heating, or an angular-momentum-rich growth history such as that suggested by \citet{hu2024}.

In this work, we construct a new sample of superthin galaxies using two-dimensional (2D) bulge/disk decomposition applied to SDSS images. Unlike the one-dimensional method used in B17, our 2D approach utilizes all pixels in the galaxy image and yields more robust structural parameters. Using multi-band imaging, SDSS $griz$ bands and UKIDSS $JHK$ bands, we examine whether the fitted disk shape of optically selected superthin galaxies is preserved in the NIR. As we demonstrate, the extreme thinness\footnote{In this work, when presenting our own measurements and related discussions, thickness-related terms such as ``thick'', ``thicker'', ``thinness'', ``thickness'', and ``thickening'' are used in a relative sense (e.g. optical versus NIR) and refer operationally to the fitted axis ratio $b/a$ (or equivalently $h_s/r_s$ for relative changes). They should not be directly equated with the intrinsic vertical scale height of the stellar disk.} of these galaxies persists in the NIR in this operational sense. Thus, our analysis directly tests the discrepancy noted by B20 and provides new insight into the wavelength dependence of disk structure in these extreme systems. Additionally, we investigate the environmental dependence of superthin galaxies using a suite of metrics spanning small to large scales (projected correlation functions, overdensity of the local environment, and large-scale structure classification), comparing our superthin sample with a control sample matched in stellar mass and colour.

This paper is organized as follows. In \secref{sec:data}, we describe the galaxy samples and the optical and NIR imaging to be used in this work. \secref{sec:shapes} explains our 2D decomposition of galaxy images and analyses of the shapes of the superthin galaxies. \secref{sec:rel_env} examines the relation with environment. We conclude with a discussion and summary in \secref{sec:summ}.

\section{Data}
\label{sec:data}

\begin{figure}
	\includegraphics[width=\columnwidth]{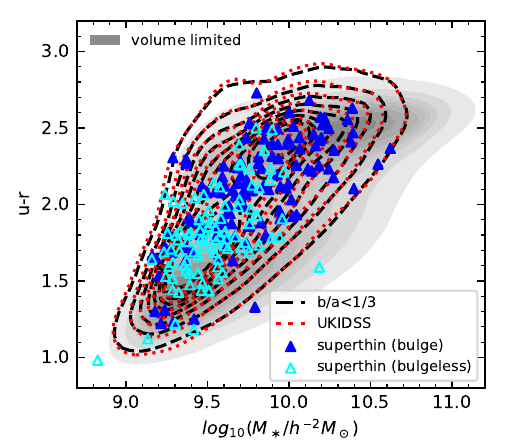}
	\caption{Distribution of galaxies in the plane of $u-r$ colour versus \lgmstar\ for the parent volume-limited sample (grey background contours), the subsample of galaxies with minor-to-major axis ratio $b/a<1/3$ (black dashed), and those with $b/a<1/3$ and UKIDSS data available (red dotted). In all cases, the contour levels are the same, enclosing 90\%, 80\%, \ldots, 10\% of a given sample. The superthin galaxies are shown as triangles: filled blue and open cyan symbols denote the bulge and ``bulgeless'' subsamples, respectively.}
	\label{fig:color_mass}
\end{figure}

\begin{figure*}
	\includegraphics[width=\textwidth]{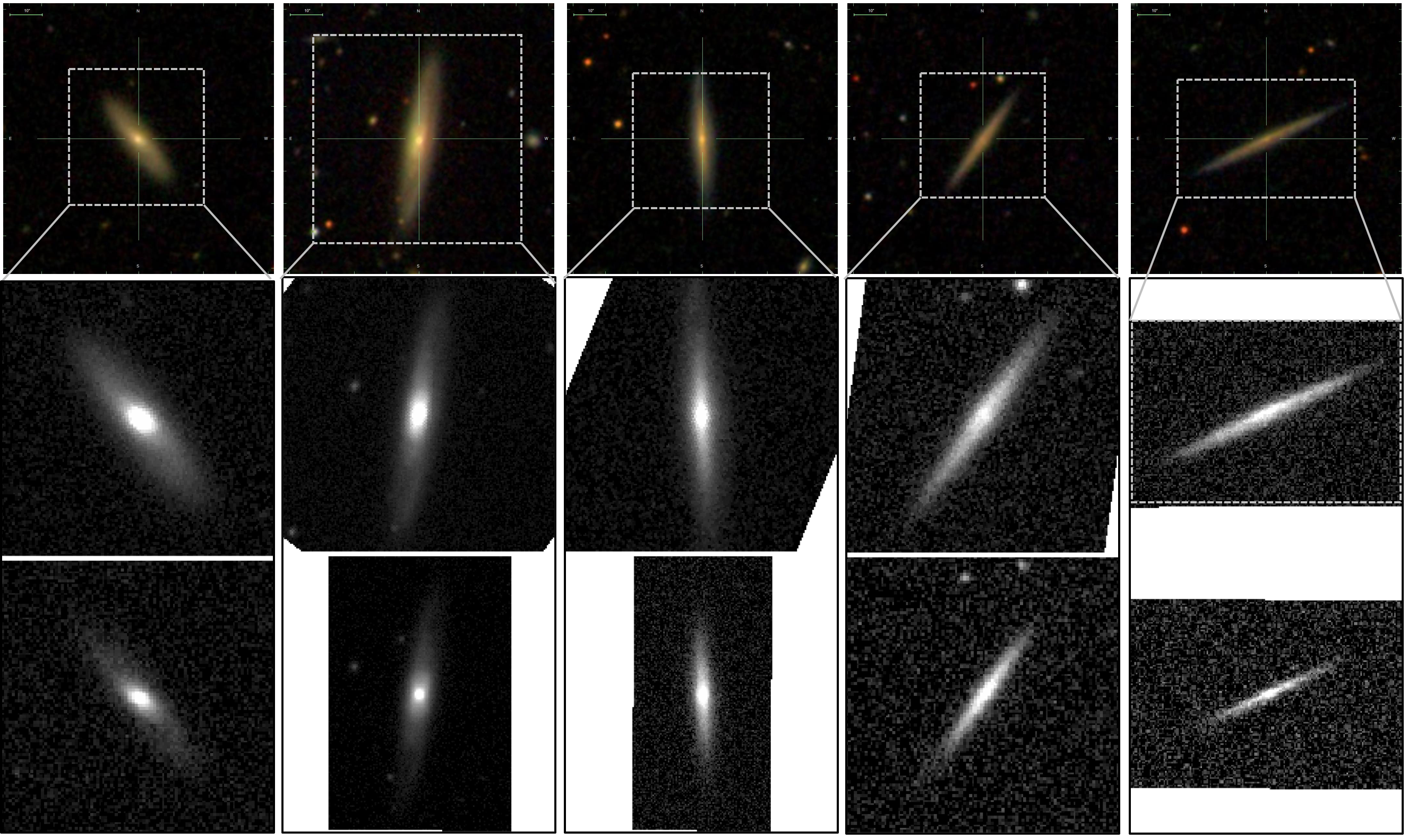}
	\caption{Examples of galaxies with different ellipticities ($1-b/a$). From left to right, the columns show galaxies with ellipticity $<0.6$, $0.6$--$0.7$, $0.7$--$0.8$, $0.8$--$0.89$, and $>0.89$, respectively. From top to bottom, the rows show the SDSS pseudo-colour image, the SDSS $r$-band image, and the UKIDSS $K$-band image.}
	\label{fig:img_demo}
\end{figure*}

\subsection{SDSS Edge-On Galaxies}
\label{sec:samp}

We select our galaxy sample from the NASA Sloan Atlas \citep[NSA;][]{blanton2011}\footnote{http://www.nsatlas.org/} which includes both spectroscopic and photometric properties for $\sim$0.6 million galaxies at $z<0.15$ from the Sloan Digital Sky Survey \citep[SDSS;][]{york2000}. For this work we restrict ourselves to relatively nearby galaxies. We select a volume-limited sample with $z$-band absolute magnitudes $M_z<-18.6$ and spectroscopically measured redshifts $0.02<z<0.05$. The absolute magnitudes are calculated from the model magnitudes from the NSA as estimated with a single S\'{e}rsic profile, corrected for Galactic extinction and $K$-corrected to redshift zero. The sample consists of $\sim$76,000 galaxies and serves as the parent sample from which we select superthin galaxies as well as control samples for comparison.

To select galaxies with a superthin disk, we further restrict the parent sample to relatively edge-on galaxies, requiring their $r$-band minor-to-major axis ratio $b/a<1/3$. The axis ratios $b/a$ are taken from the NSA, where they are obtained by fitting a two-dimensional point-spread function (PSF)-convolved S\'{e}rsic profile to the $r$-band image of each galaxy \citep{blanton2011, lupton2001}. This selection yields a sample of $\sim$15,000 galaxies with relatively (but not extremely) thin shapes.

\figref{fig:color_mass} displays the diagram of colour index $u-r$ versus \lgmstar\ for both the parent volume-limited sample (grey background contours) and the subsample of edge-on galaxies with $b/a<1/3$ (black dashed contours). The well-known colour bimodality is clearly seen in both samples. Compared to the parent sample, the edge-on galaxy sample misses the most massive galaxies with \lgmstar$\gtrsim 10.7$ and the bluest galaxies at a given mass. This can be understood because the most massive galaxies are predominantly red and elliptical, and galaxies viewed edge-on may be reddened by dust attenuation.

In \secref{sec:shapes} we perform careful bulge/disk decomposition in multiple bands for each of these galaxies. This decomposition allows us to measure the axis ratios of the disk and (if present) bulge components separately. Our final sample of superthin galaxies will then be selected based on unusually high axis ratios of the disk component, minimizing contamination from the central bulge.

\subsection{SDSS and UKIDSS imaging data}
\label{sec:image_data}

We make use of both optical images from the SDSS and near-infrared (NIR) images from the UKIRT Infrared Deep Sky Survey \citep[UKIDSS;][]{lawrence2007}. SDSS images were obtained with the Sloan 2.5-meter telescope
located at Apache Point Observatory (APO) in five bands, $ugriz$ \citep{fukugita1996, smith2002, ivezic2004}, using a large multi-CCD drift scan camera \citep{gunn1998}. Given the relatively low signal-to-noise ratio (SNR) in the $u$ band, we consider only $griz$ bands in this work, which cover wavelengths from $\sim$4000~\r{A} to $\sim$1.1~$\mu$m. The raw imaging data are sky-subtracted and calibrated, both photometrically \citep{hogg2001} and astrometrically \citep{pier2003}. Details of the SDSS data reduction pipelines and products can be found in \citet{stoughton2002}. For each galaxy in our parent sample, we retrieve the corrected image frames in FITS format directly from the SDSS Data Archive Server (DAS). Information necessary for the following analysis, such as the parameter {\tt psf\_{}width}, is fetched from the SDSS Catalog Archive Server (CAS).

UKIDSS images were obtained with the Wide Field Camera \citep[WFCAM;][]{casali2007} on the 3.8-meter United Kingdom Infra-Red Telescope (UKIRT) in five bands, $ZYJHK$ \citep{hewett2006}. We consider the $JHK$ bands, covering wavelengths from $\sim$1.1 to $\sim$2.5~$\mu$m. For each galaxy and given band, we retrieve the stacked frame from the WFCAM Science Archive \citep[WSA;][]{hambly2008}\footnote{http://wsa.roe.ac.uk/}. We note that approximately half of the galaxies in our sample are not covered by UKIDSS due to its limited sky coverage. For these galaxies, we include them in the analysis of optical images but exclude them when analysing NIR images. In \autoref{fig:color_mass}, for comparison, the distribution of the edge-on galaxies in our sample with UKIDSS data available is plotted as dotted red contours, which are almost identical to the contours of the full edge-on galaxy sample. This suggests that the former can be considered a representative subset of the latter, thus providing unbiased statistics despite the smaller sample size.

\autoref{fig:img_demo} presents five example galaxies, randomly selected from the edge-on galaxy sample but with different ellipticities defined as $\epsilon=1-b/a$. Panels from left to right display the five galaxies in order of increasing $\epsilon$, and panels from top to bottom show the SDSS $gri$ composite image, the SDSS $r$-band image, and the $K$-band image from UKIDSS, respectively. Although the number of galaxies is small, the figure reveals two interesting trends. First, as ellipticity increases, the bulge component becomes less and less pronounced. In the two right-most galaxies with the highest ellipticities, the bulge is no longer apparent. Second, although these relatively thin galaxies are selected in the optical, their NIR images show a similarly thin shape, but with smaller radii in both the major and minor axes. In what follows we use our sample and the SDSS/UKIDSS images to perform statistical and quantitative investigations of these trends.

\subsection{Reference and Random Samples}

In \secref{sec:rel_env} we investigate the clustering of our superthin galaxies by estimating the projected two-point cross-correlation function of the superthin galaxy sample with respect to a reference sample, with the help of a random sample to account for the selection effects in the reference sample. Following \citet{li2006a, li2006b}, we construct the reference and random samples from the New York University Value-Added Galaxy Catalog (NYU-VAGC)\footnote{\url{http://sdss.physics.nyu.edu/vagc/}}, which was compiled by \citet{blanton2005} from the seventh data release of the SDSS \citep[DR7;][]{SDSS-DR7}. The reference sample consists of approximately half a million galaxies with spectroscopically measured redshifts $0.01<z<0.2$ and $r$-band apparent Petrosian magnitude $r<17.6$. The random sample is constructed to have the same selection effects as the reference sample, but with a randomised spatial distribution and a sample size ten times larger than that of the reference sample.

\subsection{Overdensity and Large-Scale Structure Classification} \label{subsec:dens_lss}

In \secref{sec:rel_env} we also investigate the local environment of the superthin galaxies by considering two quantities: the density contrast ($\rho/\bar{\rho}$) and the type of the large-scale structure (LSS) of the local environment. We take the measurements of $\rho/\bar{\rho}$ and LSS type from the SDSS density field reconstruction of \citet{Wang2016ELUCID}, which fully covers the volume of our galaxy sample. The density field was obtained by running a constrained $N$-body simulation starting from the initial density field of the nearby universe, reconstructed to closely reproduce the distribution of galaxy groups in the SDSS DR7. The reconstructed density field is then smoothed on different scales, and the resulting cosmic velocity and tidal fields are used to classify each grid cell in the smoothed field into one of four categories: void, filament, sheet, or cluster. In this work we consider density fields smoothed on scales of 1, 2, and $3\,h^{-1}\,\text{Mpc}$, whereas the $2\,h^{-1}\,\text{Mpc}$ version is adopted for LSS type classification.

\begin{figure*}
    \includegraphics[width=\textwidth]{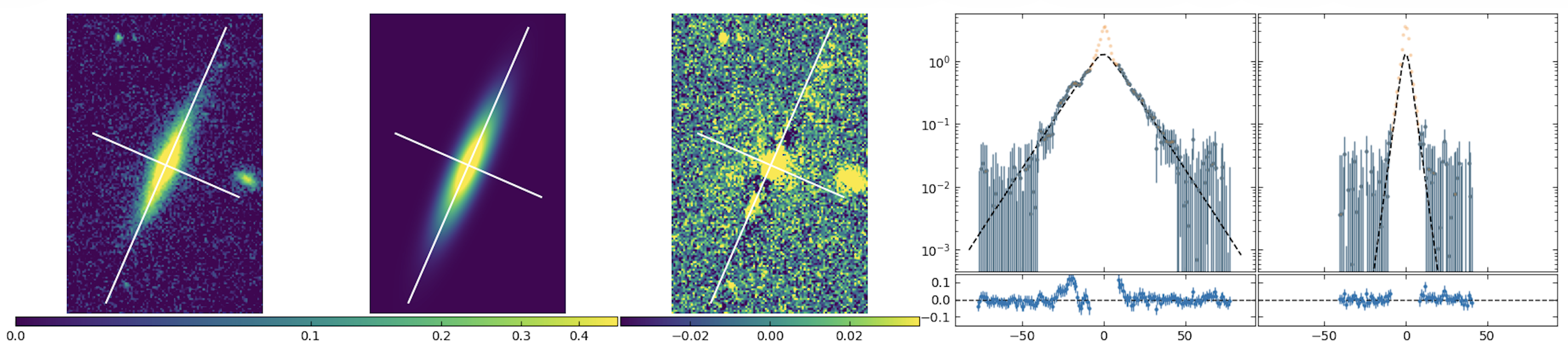}
    \includegraphics[width=\textwidth]{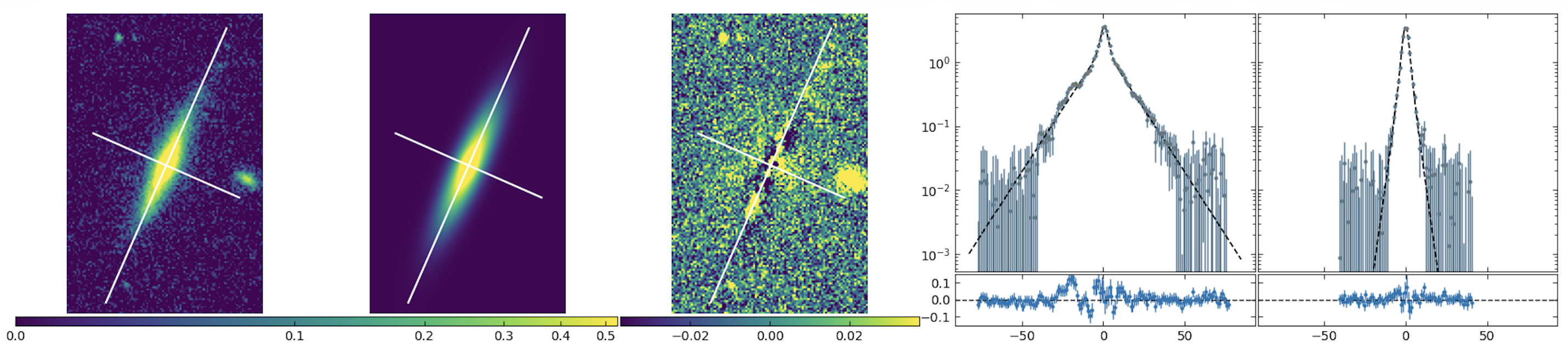}
    \caption{Demonstration of the $b/a$ measurement. The upper row shows the disk-only fit with the central region masked, and the lower row shows the final bulge+disk fit. In each row, the three left panels show the observed image, model, and residual, and the two right panels show the profiles along the major and minor axes, indicated by white lines in the images. In the lower-row profile panels, the masked central pixels are shown in a lighter colour.}
    \label{fig:ellip_galfit}
\end{figure*}

\begin{figure*}
    \includegraphics[width=\textwidth]{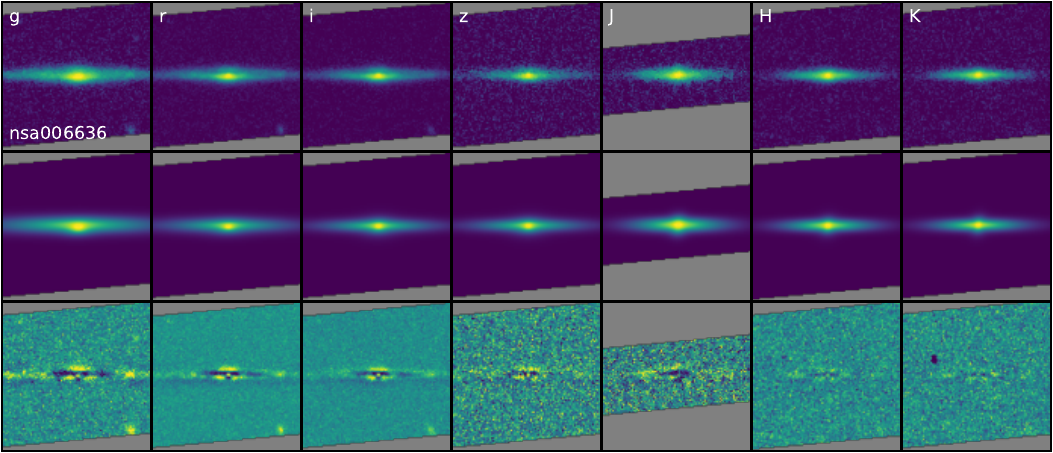}
    \includegraphics[width=\textwidth]{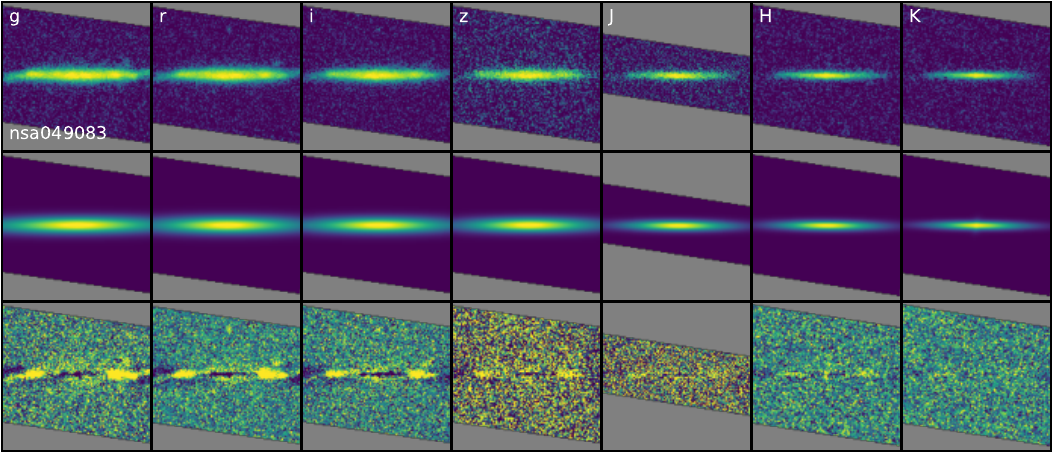}
    \caption{Demonstrations of two superthin galaxies across multiple bands (NSA IDs are marked in the first panel of each example). Columns from left to right correspond to the SDSS $g$, $r$, $i$, $z$ bands and the UKIDSS $J$, $H$, $K$ bands. For each galaxy, three rows show the observed image, model image, and residual image from top to bottom. The disk component is fitted with
    an exponential disk profile,
    with or without a {\sersic} bulge for the upper and lower examples, respectively. Images in different bands are resampled to the same region.}
	\label{fig:demos_bands}
\end{figure*}

\section{Shapes of superthin galaxies}
\label{sec:shapes}

\subsection{Image decomposition and modelling}
\label{sec:decomp}

We use the widely adopted software \texttt{GALFIT}\footnote{\url{https://users.obs.carnegiescience.edu/peng/work/galfit/galfit.html}} \citep{peng2002, peng2010} for image decomposition. GALFIT determines the best-fitting model parameters by minimizing
\begin{equation}
\chi^2 = \sum_{\mathrm{mask}(x_i, y_i)=0} \frac{\left[I_{obs}(x_i, y_i)-(I_{mod} \circledast \mathrm{PSF})(x_i, y_i)\right]^2}{\sigma(x_i, y_i)^2},
\label{eq:chisq}
\end{equation}
where $I_{obs}(x_i, y_i)$ is the observed image, $\mathrm{mask}(x_i, y_i)$ is the mask image with zero/non-zero values for good/bad pixels, $\sigma(x_i, y_i)$ is the sigma image, and $I_{mod} \circledast \mathrm{PSF}$ is the convolution of the model image with the instrumental point-spread function (PSF). The bad pixels include, for example, saturated pixels and pixels belonging to foreground or background sources. For a single galaxy image, the model is written as $I_{mod}=I_{gal}+I_{sky}$, where $I_{gal}$ and $I_{sky}$ are the galaxy and sky models, respectively. In this work, we fix $I_{sky}$ to a pre-measured constant over the image, while $I_{gal}=I_{gal}(x,y;\boldsymbol{\theta})$ is the 2D spatial profile whose parameters $\boldsymbol{\theta}$ are determined by the $\chi^2$ minimization.

Before fitting, we prepare the required image products for each galaxy and band. We first use \texttt{SExtractor}\footnote{https://www.astromatic.net/software/sextractor} \citep{bertin1996} to identify the target galaxy and other sources, including foreground stars and background galaxies, adopting a threshold of $2\sigma$ for each pixel. We then measure the basic photometric parameters of the target galaxy, including its centre, major-axis position angle, axis ratio, and effective radius. We also obtain the noise image (the $\sigma$ map) and the PSF image. The $\sigma$ image is computed by
\begin{equation}
\sigma = \sqrt{(I_{obs} - I_{\mathrm{sky}})/\mathrm{GAIN} + \sigma _{\mathrm{sky}}^2},
\label{eq:sigma}
\end{equation}
where $I_{obs}$ is the intensity of a given pixel, $I_{\mathrm{sky}}$ and $\sigma _{\mathrm{sky}}$ are the mean and RMS of the sky background, and the GAIN factor is fetched from the SDSS data server or the UKIDSS image FITS header. The PSF image is produced using a Gaussian profile with the FWHM fetched in the same way. Additional technical details about the PSF and sky level are described in \appref{app:tech_decomp}.

\begin{figure*}
	\centering
	\includegraphics[width=\textwidth]{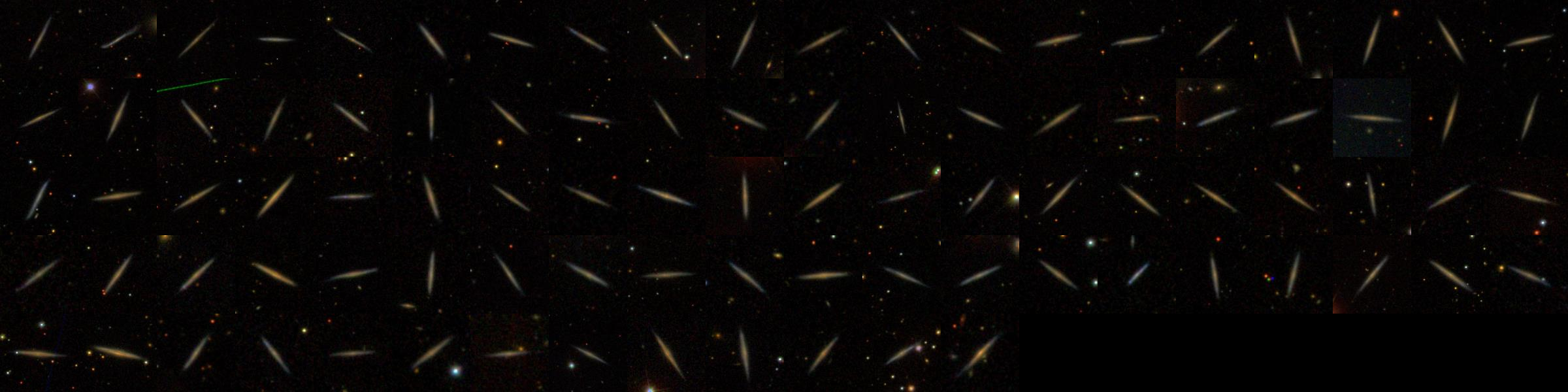}\\
	\vspace{0.1mm}
	\includegraphics[width=\textwidth]{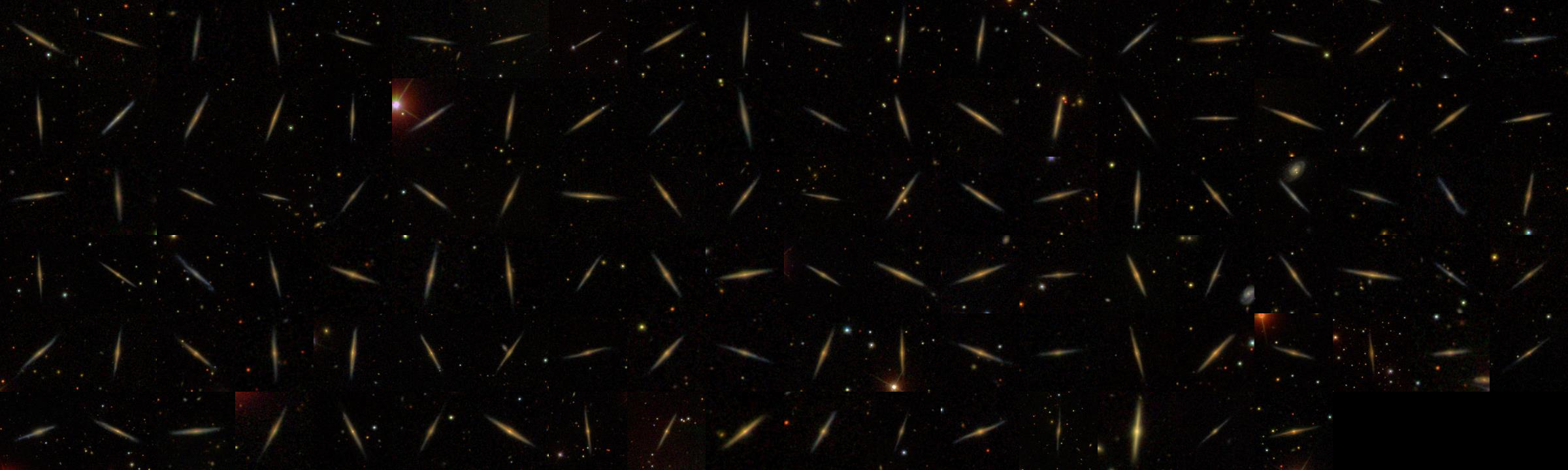}\\
	\caption{Superthin galaxy sample. The SDSS pseudo-colour images are divided into two subgroups according to the fitted bulge component: the ``bulgeless'' subsample is shown in the upper panel, and the bulge subsample is shown in the lower panel. The two subgroups contain 93 and 117 galaxies, respectively. Galaxies in the upper panel are ordered by increasing disk $b/a$, while those in the lower panel are ordered by increasing $B/T$.}
	\label{fig:img_sup_bins}
\end{figure*}

\begin{figure}
\centering
	\includegraphics[width=\columnwidth]{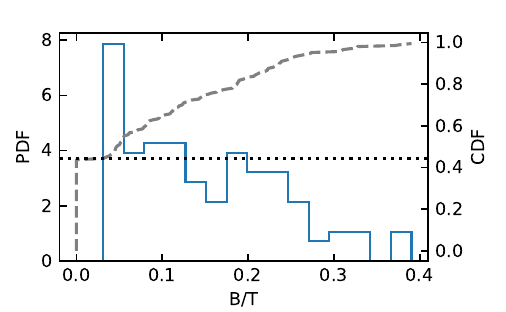}
	\caption{Differential (histogram) and cumulative (dashed line) distributions of the $r$-band bulge-to-total luminosity ratio ($B/T$) of the superthin galaxy sample. The horizontal dotted line indicates the fraction of bulgeless galaxies, $\sim 44\%$.}
	\label{fig:hist_bt}
\end{figure}

  For the galaxy model, we decompose it into two components: a disk plus, if present, a bulge. Following common practice, both disk and bulge are assumed to be axisymmetric, and their radial surface brightness distributions are modelled by a S\'ersic profile \citep{sersic1968, ciotti1991}:

\begin{equation}
I_{gal}(r) = I_e \exp\left\{-b_n \left[\left(\frac{r}{r_e}\right)^{1/n}-1\right]\right\},
\label{eq:sersic}
\end{equation}

where $r_e$ is the effective radius, and $I_e$ is the surface brightness at $r_e$, i.e. $I_e=I(r_e)$. The normalization constant $b_n = 0.868 n - 0.142$ \citep{caon1993}, so that the effective radius $r_e$ encloses half of the total light in the given band. The S\'ersic index $n$ characterizes the concentration of the light distribution. For instance, an index of $n=1$ corresponds to an exponential disk \citep{freeman1970}, while $n=4$ corresponds to the $R^{1/4}$ law which was conventionally used to describe classical bulges and ellipticals \citep{devau1948}. Due to inclination, the galaxy is not circular at a given radius but is close to elliptical in shape. For a pixel in the observed image with coordinates $(x,y)$, the corresponding radius in the model profile is given by
\begin{equation}
r = \sqrt{(x-x_c)^2 + \left(\frac{y-y_c}{b/a}\right)^2},
\label{eq:ellip}
\end{equation}
where $x_c$ and $y_c$ are the coordinates of the galactic centre, and $a$ and $b$ are the major- and minor-axis lengths of the image. The parameter $r_e$ in \eqref{eq:sersic} can be used as a characteristic value of $a$, so that only the axis ratio $b/a$ needs to be fit as a single parameter. For a given component (bulge or disk in our case), all the model parameters ($n$, $r_e$, and $b/a$) are assumed to be independent of radius; that is, they are constant across the whole galaxy. To start the fitting, initial guesses for the model parameters are required. In this work, they are determined based on the \sex{} results. In detail, for the galactic center $(x_c, y_c)$ and the position angle, both components initially share the same starting values, taken directly from \sex{}. For the effective radius $r_e$ and the axis ratio $b/a$, the disk values are also taken directly, while for the bulge they are set to $0.8\,r_e$ and 0.8, respectively. The \sersic{} index $n$ of the bulge (not needed for the disk, where it is fixed to 1) is initially set to 2. During the fitting, all these parameters are left free and are finally determined by $\chi^2$-minimization as above.

In cases where the image does not present a significant
bulge component in the centre, the whole image is fitted
with a single exponential disk component, which is modelled
by a S\'ersic profile with an index of $n=1$.
For images with a significant bulge,
the model parameters of the disk and bulge components can be
determined simultaneously by fitting the model profiles
at all radii in
\eqref{eq:sersic}
to all pixels in the
image, from the galactic centre out to a substantially large outer radius.
By doing so, however, one may suffer from strong degeneracy
between the two model components. To overcome this problem,
we opt to fit the disk and bulge components separately.
First, we mask out the central part and fit the
outer part of the image with an exponential disk as
modelled by a S\'ersic profile with an index of $n=1$.
Similarly, we mask out the outer part and apply a S\'ersic model
with $n>1$ to fit the central bulge.
Next, we use the model parameters determined separately
as initial guesses and fit the whole image simultaneously with a
combination of the two model components. In this step
we fit the model parameters for one component only,
while fixing the model parameters for the other component.
This process is iteratively repeated to fit the two components
in turn, ending with stable model parameters for both
components (typically after three rounds of fitting).
The mask for the central part is adopted to be a circle
with radius $r_e$ centred on the galactic centre,
with both $r_e$ and the galactic centre determined by
\texttt{SExtractor} in the first step. We adopt a circular mask rather than an elliptical one because the purpose of the mask is simply to cover any possible bulge in the central region. A circular mask is independent of inclination and the apparent bulge shape, avoiding assumptions about the intrinsic geometry. The key parameter is therefore the mask size. We have tested with a range
of mask radii from $0.5$ to $1.5\,r_e$, and we find that our
results are robust to the choice of this radius.

As an example, \autoref{fig:ellip_galfit} shows the fitting
result for the $r$-band image of one of our galaxies.
The upper panels show the result for the model with a disk
component only, for which the central part is masked out
during the fitting. The lower panels display the result
of the final best-fit model which includes both the disk and
the bulge components. In each row the panels from left
to right display the observed image, the model image,
the residual in the model-subtracted image, and the
surface brightness profile along the major and minor axes,
respectively.

\figref{fig:demos_bands} displays the observed image, the best-fit S\'{e}rsic model, and the model-subtracted residual image for two example superthin galaxies (one with a central bulge and one without), across multiple bands, from SDSS $g$, $r$, $i$, $z$ to UKIDSS $J$, $H$, $K$. For each galaxy, images in different bands are rotated and resampled to the same sky region, aligned with respect to the disk plane in the $r$ band. These images offer a direct visual impression of the galaxy shape across bands. A simple inspection reveals that these superthin galaxies retain their superthin morphology from the optical to the NIR.

 \begin{figure*}
    \includegraphics[width=\textwidth]{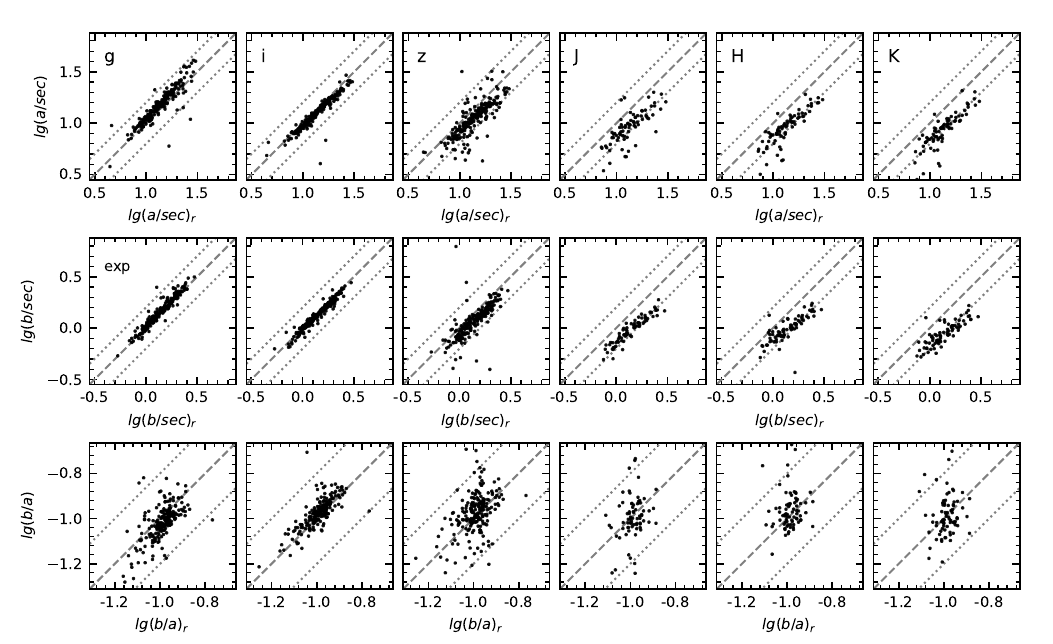}
    \caption{Comparison of disk shape parameters in other bands with those in the $r$ band. Three rows from top to bottom correspond to $a$, $b$, and $b/a$ of the exponential disk model, all in logarithmic scale. The x-axis shows parameters in the $r$ band, and the y-axis shows parameters in other bands: SDSS $g$, $i$, $z$, and UKIDSS $J$, $H$, $K$. The dashed diagonal line in each panel indicates equal comparison, and the two dotted lines indicate $\pm$0.2 dex.}
	\label{fig:comp_ba_bands}
\end{figure*}

We do not attempt a separate thin+thick disk decomposition. The depth and spatial resolution of the present images make it difficult to resolve such subtle vertical substructure robustly, and adding a second disk component would introduce strong degeneracies between components whose physical interpretation would be uncertain. The small residuals in Figs. \ref{fig:ellip_galfit} and \ref{fig:demos_bands} indicate that the adopted bulge+disk model is sufficient for measuring the overall disk shape in the present data. We therefore use the single fitted disk component to test the relative waveband dependence of the disk shape. A thick disk could still be present, but our analysis shows that any such component is not prominent enough, relative to the thin disk and within the available image depth, to change the overall fitted disk axis ratio.

\subsection{Selection of superthin galaxies}
\label{sec:superthin_sample}

We select superthin galaxies based on the best-fit model in the $r$ band. Specifically, a galaxy is classified as a superthin galaxy if its exponential disk has a minor-to-major axis ratio $b/a<1/9$, equivalent to $a/b>9$. This follows the classical criterion introduced by \citet{goad1979, goad1981} and widely adopted in subsequent studies.

This selection yields a sample of 343
superthin galaxies, $\sim 0.45\%$ of the parent volume-limited
sample (see \S~\ref{sec:samp}). We then visually examine
the SDSS $gri$ composite image of each galaxy and exclude
37 galaxies with apparent problems in the image data.
Among the remaining galaxies, we find that about 30\% show prominent dust lanes or clear residual structures after subtraction of the best-fit model. Given the potential impact of these features on shape measurements, we exclude these galaxies through visual inspection of the residual images. This leaves a final sample of 210 galaxies, which are shown as blue or cyan triangles in \autoref{fig:color_mass}.
 An example of the excluded galaxies is provided in \appref{app:dust}, along with discussions of their distribution and their impact on the subsequent analysis.
When compared to the parent sample of edge-on galaxies with $b/a<1/3$,
the superthin galaxy sample is preferentially located in the middle colour regime, thus lacking both the reddest and the bluest galaxies at a fixed mass. The lack of bluest galaxies might be caused by dust reddening as mentioned above, while the lack of reddest galaxies may indicate that low-level star formation is still present in superthin galaxies.

\autoref{fig:img_sup_bins} displays
the SDSS composite images of all 210 superthin galaxies, which are divided into two categories: 93 ``bulgeless'' galaxies
fitted with a disk component only, and 117 galaxies
fitted with both disk and bulge components. In the figure,
the galaxies in the ``bulgeless'' category are ordered by
increasing $b/a$, while those in the bulge category are
ordered by increasing the $r$-band bulge-to-total luminosity
ratio $B/T$.
These two subgroups are also shown as cyan and blue triangles, respectively, in \figref{fig:color_mass}. The ``bulgeless'' category is mostly located in the blue cloud, while the bulge category spans the red sequence and green valley.
\autoref{fig:hist_bt} shows the distribution of $B/T$ 
for the superthin galaxies. Most superthin
galaxies have relatively weak bulges: $\sim44\%$ have
no bulge, and $84\%$ have $B/T < 0.2$.
 \autoref{tab:sample_stats} summarises the median values and $[0.2, 0.8]$ quantile ranges of some key parameters for the total superthin sample, and the bulge/``bulgeless'' subsamples.

\begin{table*}
\centering
\caption{Median values and $[0.2, 0.8]$ quantile ranges of some key parameters for the total superthin sample, and the bulge/``bulgeless'' subsamples. From left to right, the columns are: (1) Superthin galaxy sample; (2)-(3) Stellar mass and colour index $u-r$, shown in \figref{fig:color_mass}; (4) Bulge-to-total ratio, shown in \figref{fig:hist_bt}; (5)-(6) Redshift and $z$-band absolute magnitude, used for the parent volume-limited sample selection (\secref{sec:samp}).}
\label{tab:sample_stats}
\begin{tabular}{c|cc|c|cc}
    \hline
    Sample & $\log_{10}(M_\ast/h^{-2}M_\odot)$ & $u-r$ & $B/T$ & Redshift & $M_z$ \\
    \hline
    Total & 9.63 $[9.39, 9.94]$ & 1.93 $[1.69, 2.26]$ & 0.046 $[0, 0.19]$ & 0.037 $[0.028, 0.045]$ & -19.7 $[-20.3, -19.0]$ \\
    Bulge & 9.76 $[9.48, 10.06]$ & 2.12 $[1.81, 2.40]$ & 0.13 $[0.057, 0.23]$ & 0.036 $[0.028, 0.044]$ & -19.8 $[-20.6, -19.2]$ \\
    Bulgeless & 9.51 $[9.34, 9.74]$ & 1.78 $[1.59, 2.02]$ & 0 $[0, 0]$ & 0.038 $[0.028, 0.046]$ & -19.3 $[-19.8, -18.9]$ \\
    \hline
\end{tabular}
\end{table*}

For comparison with previous studies,
for each superthin galaxy we have also fitted the images
in different bands using the following three-dimensional model which is
frequently considered in previous studies for edge-on
galaxies \citep{kruit1981}:
\begin{equation}
\rho(r, z) = \rho_0 \exp\left(-\frac{r}{r_s}\right) \sech^2\left(\frac{z}{h_s}\right),
\label{eq:disk3d}
\end{equation}
where $r_s$ and $h_s$ are the scale length and scale height,
respectively. In the edge-on view, the corresponding two-dimensional
surface brightness distribution is given by integrating the
above equation along the line of sight:
\begin{equation}
I(R, z) = I_0 \left(\frac{r}{r_s}\right) K_1\left(\frac{r}{r_s}\right) \sech^2\left(\frac{z}{h_s}\right).
\label{eq:edgeon}
\end{equation}
Here, $K_1(x)$ is the modified Bessel function of the second kind.
The ratio $(h_s/2)/r_s$ approximates the axis ratio $b/a$
obtained above from fitting the isophotes with the S\'ersic model. The measured parameters will be analysed in the following subsection.


\subsection{Disk thinness according to waveband}
\label{sec:thin_band}

\begin{figure}
    \centering
    \includegraphics[width=0.7\columnwidth]{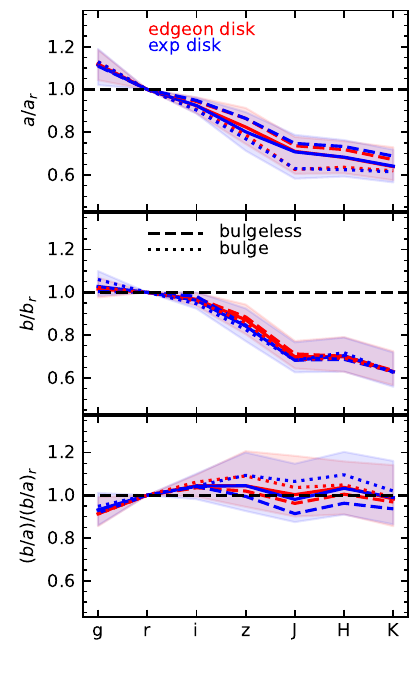}
    \caption{Disk shape parameter ratios as a function of waveband, normalized to the $r$ band. The x-axis in each panel gives the band, and the y-axis gives the ratio relative to the $r$-band value. From top to bottom, the panels correspond to $a$ (or $r_s$), $b$ (or $h_s$), and $b/a$ (or $h_s/r_s$). Different colours indicate different disk models: red for the edge-on disk and blue for the exponential disk. Line styles distinguish samples: solid for the full superthin sample, dashed for the bulgeless subsample, and dotted for the bulge subsample. The shaded region around the solid line represents the [0.2, 0.8] quantile range of the full superthin sample, and each line shows the median of the corresponding (sub)sample.}
    \label{fig:ba_bands}
\end{figure}

\figref{fig:comp_ba_bands} compares the disk shape parameters measured in different bands, using the $r$ band as a reference. For simplicity, we present the results only for the exponential disk model. The three rows correspond, from top to bottom, to the major-axis length $a$, the minor-axis length $b$, and their ratio $b/a$. The dashed line indicates the $1:1$ relation, while the two dotted lines mark a $\pm0.2$ dex difference. The plots show that both $a$ and $b$ decrease from shorter to longer wavelengths; at the $K$ band, they are roughly $0.2$ dex smaller than in the $r$ band. However, the axis ratio $b/a$, which quantifies the disk thickness, remains approximately constant, with a scatter significantly smaller than $0.2$ dex. We have also compared the scale lengths ($r_s$, $h_s$) and ratio ($h_s/r_s$) measured from the edge-on disk model, finding very similar results.

These trends are more clearly seen in \figref{fig:ba_bands}, which shows the band-dependent variation of disk shape parameters relative to the $r$ band. From top to bottom, the panels correspond to $a$, $b$, $b/a$ for the exponential disk model (\eqref{eq:sersic} and \eqref{eq:ellip}), or $r_s$, $h_s$, $h_s/r_s$ for the edge-on disk model (see \eqref{eq:edgeon}). Different disk models are distinguished by colour: red for the edge-on disk and blue for the exponential disk. The solid line represents the whole superthin sample, while the dashed and dotted lines denote the bulgeless and bulge subsamples, respectively. The shaded regions indicate the $[0.2, 0.8]$ quantile range of the full sample. As seen, the scale lengths along both the major axis ($a$ or $r_s$) and the minor axis ($b$ or $h_s$) decrease from optical to NIR bands. Both decreases follow nearly the same pattern, reaching a ratio of $\sim0.6$ between the $K$ and $r$ bands. This results in an almost invariant axis ratio ($b/a$ or $h_s/r_s$), varying by $\lesssim10\%$ across all bands and for both models. Comparing the bulgeless and bulge subsamples reveals no significant offset; both follow the same trend as the whole sample, indicating that the presence of a bulge does not affect the disk shape of the superthin galaxies.

 The conclusion is also insensitive to the exact superthin threshold. In \figref{fig:comp_ba_bands}, adopting thresholds such as $b/a<1/8$ or $b/a<1/10$ simply changes the cutoff along the x-axis of the bottom row; galaxies on either side of these nearby cutoffs show the same weak waveband dependence of $b/a$. We also repeated the median band-dependence test shown in \figref{fig:ba_bands} with these alternative thresholds and found no significant change in the result.

\subsection{Comparison with previous studies}
\label{subsec:comp_bizyaev}

\begin{figure}
    \centering
    \includegraphics[width=\columnwidth]{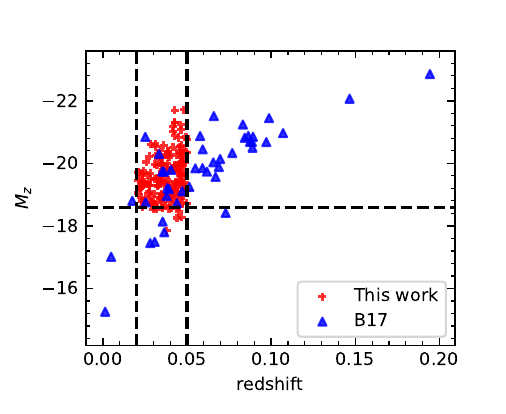}
    \caption{Distribution of the superthin galaxy samples from B17 (blue triangles) and the present work (red crosses) in the diagram of $z$-band absolute magnitude versus redshift. The horizontal and vertical dashed lines indicate the sample selection criteria of the present work.}
    \label{fig:dist_samps_mz_z}
\end{figure}

\begin{figure}
    \centering
    \includegraphics[width=\columnwidth]{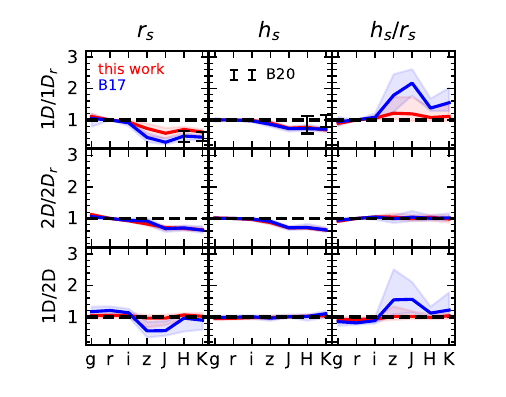}
    \caption{Comparison of the 1D and 2D measurements for our sample (red) and the B17 sample (blue). The three columns correspond to $r_s$, $h_s$, and $h_s/r_s$. The upper two rows show band-to-$r$ ratios measured with the 1D and 2D methods, respectively, and the bottom row shows the ratio between the 1D and 2D measurements. Lines and shaded regions show the median and [0.2, 0.8] quantile range, as in \figref{fig:ba_bands}. The error bars in the upper row show the B20 measurements in the $H$ and $K$ bands relative to the $r$ band.}
    \label{fig:comp_1d_2d}
\end{figure}

\begin{figure*}
    \includegraphics[width=\textwidth]{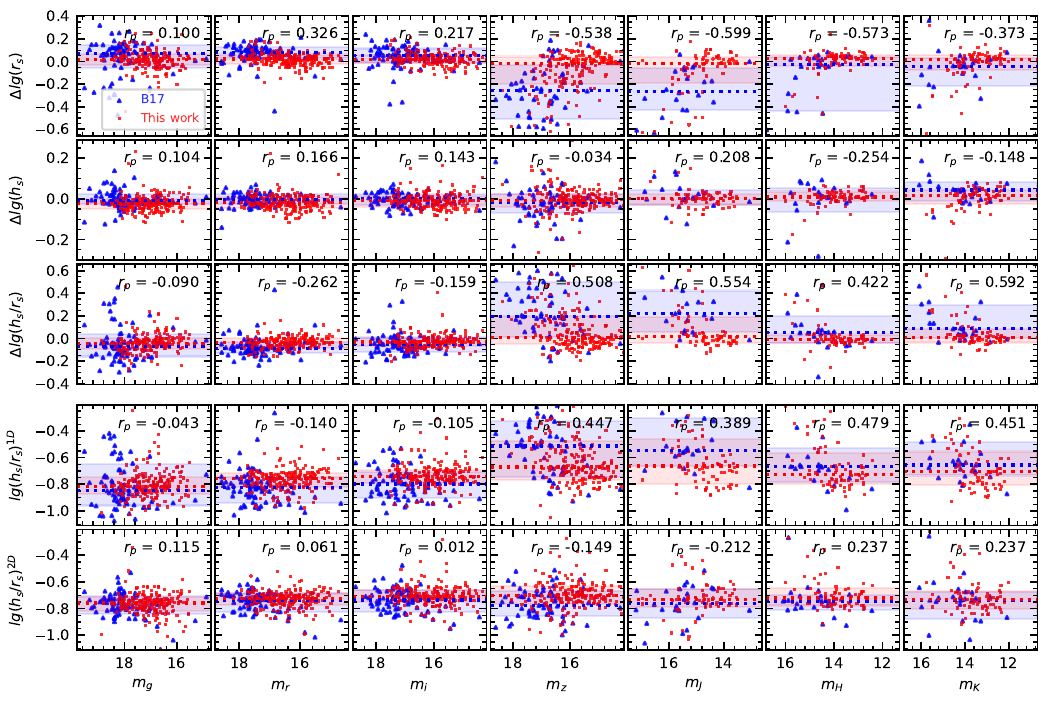}
    \caption{Disk shape parameters from the 1D and 2D methods as a function of galaxy apparent magnitude. Columns from left to right correspond to the SDSS $griz$ bands and the UKIDSS $JHK$ bands, respectively. The top three rows show the logarithmic difference between the two methods ($\log_{10}(1D/2D)$ for $r_s$, $h_s$, and $h_s/r_s$ of the edge-on disk model; \eqref{eq:disk3d}). The bottom two rows show $\log_{10}(h_s/r_s)$ from the 1D and 2D methods, respectively. Red and blue colours correspond to our sample and the B17 sample, respectively, with crosses and triangles marking individual galaxies. Dotted horizontal lines and shaded regions mark the median and the [0.2, 0.8] quantile range of the y-axis parameters. The Pearson correlation coefficient $r_p$ calculated for all galaxies is given in the upper right of each panel.}
    \label{fig:comp_rs_mag_r}
\end{figure*}

Our results presented above clearly show that optically identified superthin galaxies remain superthin in the NIR. We find no evidence of waveband-dependent variation in disk shape, nor of a NIR-dominant vertically extended component prominent enough to alter the fitted global disk shape. This result appears to disagree with the previous study by B20, who also examined the NIR shape of superthin galaxies selected from SDSS but found that the disks become thicker (i.e., exhibit larger minor-to-major scale length ratios) in the NIR than in the optical. Their disk model is identical to our edge-on disk (\eqref{eq:disk3d}). They found that in the NIR, the disk scale length (along the major axis) is significantly smaller than in the optical, while the scale height (along the minor axis) decreases less markedly, resulting in a notably rounder disk shape in the NIR.

The discrepancy between B20 and the present work could arise from differences in the samples used and/or the methods employed to measure disk shapes. The comparison sample from B17 consists of 85 superthin galaxies selected from SDSS based on a disk scale length ratio $r_s/h_s > 9$ in the $r$ band, and 49 of these galaxies were followed up in the NIR by B20. This selection approximates an isophotal axis ratio $r_s/(h_s/2) > 18$ at large radius --- a stricter criterion than that used in our work and in previous studies (e.g., \citealt{goad1981}). Furthermore, unlike the volume-limited sample used in our work, the B17 sample is flux-limited, thus spanning a relatively broad range in both redshift and luminosity (see \figref{fig:dist_samps_mz_z}). Those authors adopted a different approach to measure disk shapes, fitting one-dimensional (1D) profiles independently along the major and minor axes (see \citealt{bizyaev2014}), in contrast to the two-dimensional (2D) fitting approach employed in our study. To investigate the origin of the discrepancy, we have applied both the 1D method described in B17 and our 2D method to the B17 comparison sample as well as to our own sample.

\begin{figure*}
    \includegraphics[width=\textwidth]{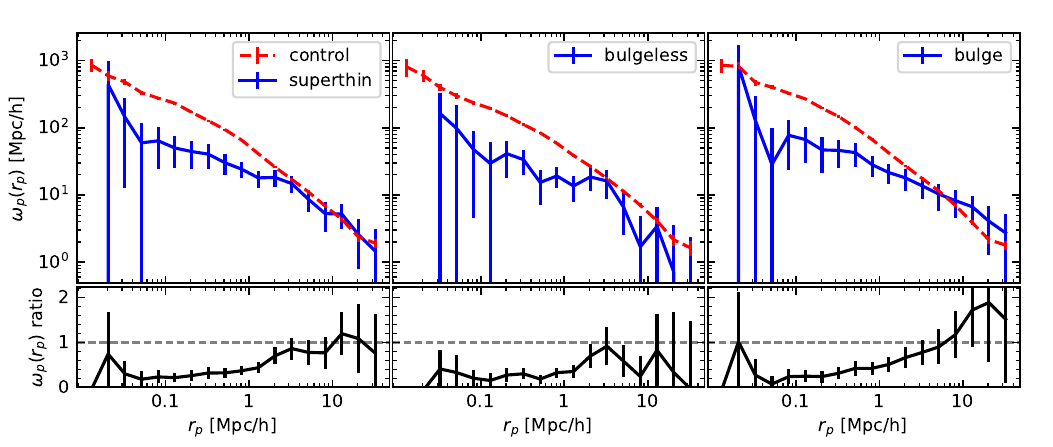}
    \caption{Projected cross-correlation functions for the superthin samples (blue solid lines) and their matched control samples (red dashed lines). The upper panels show the projected cross-correlation function $w_p(r_p)$ and the lower panels show the $w_p(r_p)$ ratio relative to the control sample. From left to right, the panels show the full superthin sample, the ``bulgeless'' subsample, and the bulge subsample.}
        \label{fig:cross_2bins}
\end{figure*}

\figref{fig:comp_1d_2d} summarises the results of applying the two measurement methods to the two superthin galaxy samples. In each panel, the red and blue lines distinguish the two samples (our sample in red and the B17 sample in blue). The columns, from left to right, correspond to the parameters $r_s$, $h_s$, and $h_s/r_s$ of the edge-on disk model (\eqref{eq:disk3d}). The top two rows show the results of the 1D and 2D methods, with the $y$-axis giving the ratio of a measurement in a given band to that in the $r$ band; the bottom row plots the ratio between the two methods. The measurements reported in B20 are also overlaid for comparison, plotted as error bars in the top row for the $H$ and $K$ bands. As can be seen from the top panels, when the 1D method is adopted, the two samples yield different results: the B17 sample essentially recovers the B20 conclusion --- compared to the optical bands, the NIR bands exhibit smaller $r_s$, similar $h_s$, and thus significantly larger $h_s/r_s$, while our sample leads to the same results as obtained above with the 2D method. The panels in the middle row show that, when the 2D method is applied, the two samples agree well with each other: the disk axis ratio is invariant from the optical to the NIR, although the scale lengths along both the major and minor axes decrease with increasing wavelength. The bottom panel reinforces the fact that whether the two methods agree or disagree depends on the sample used. This result suggests that the discrepancy between B20 and the present work is caused by both the different methods and the different samples.

In light of this, we have examined how the method discrepancy depends on galaxy properties. \figref{fig:comp_rs_mag_r} illustrates the effect of apparent magnitude, which largely explains the discrepancy between B20 and our work. The upper three rows show the differences in $r_s$, $h_s$, and $h_s/r_s$ between the 1D and 2D methods as a function of apparent magnitude, for bands from $g$ to $K$. Results for the B17 sample (blue) and our sample (red) are shown separately. The two methods yield consistent $h_s$ measurements at all magnitudes. For $r_s$, however, the 1D method overestimates it for faint optical galaxies ($\Delta\log r_s\sim0.1$ dex at $m\sim20$) but underestimates it for bright ones ($\sim-0.1$ dex at $m\sim14$); in the NIR, it strongly underestimates $r_s$ for faint galaxies ($\lesssim-0.4$ dex). Consequently, for faint galaxies, the 1D method slightly underestimates $h_s/r_s$ in the optical but significantly overestimates it in the NIR. This explains the B20 result: their B17 sample, dominated by faint galaxies, yields an artificially increasing $h_s/r_s$ with wavelength. The lower two rows show $h_s/r_s$ measurements from each method separately. The 2D method results are independent of magnitude and band, while the 1D method shows clear magnitude- and band-dependent biases.

The magnitude-dependent bias of the 1D method is readily understood. This method is applied to regions of a galaxy image above $2\sigma$ of the background, defined by an ellipse. Apparent magnitude affects the size of these regions and thus influences the 1D measurements. The fainter apparent magnitudes of the B17 sample arise directly from its broader redshift range. Because the 1D method uses only a subset of pixels, it is more sensitive to noise than the 2D method. We therefore conclude that our results are more reliable, given both the robustness of the 2D method and the volume-limited nature of our sample.

\section{Clustering and Environment}
\label{sec:rel_env}

We now examine whether superthin galaxies occupy environments that differ from those of otherwise similar galaxies. We use three complementary diagnostics: the projected cross-correlation function, the local overdensity, and the large-scale-structure (LSS) classification. The reference and random samples used for the correlation analysis are described in Section~2.3, and the overdensity and LSS measurements are described in Section~2.4; here we focus on the comparison between the superthin samples and matched control samples.

\subsection{Control samples}
\label{sec:cntr}

Galaxy environment depends strongly on stellar mass and colour, so we compare each superthin sample with a control sample matched in these quantities. The controls are drawn from the parent volume-limited sample after excluding the corresponding superthin galaxies. We use a kernel-based resampling procedure to match the distributions of \mstar\ and $g-r$ colour for the full superthin sample, and separately for the ``bulgeless'' and bulge subsamples. For the LSS comparison, we additionally match the local overdensity smoothed on $1\,h^{-1}\,\mathrm{Mpc}$, so that the LSS test probes any residual dependence on cosmic-web type at fixed stellar mass, colour, and smaller-scale density. For measurements involving the ELUCID density field, the control samples are constructed only from galaxies with available overdensity and LSS measurements. All matchings are performed on the joint distribution of the corresponding control parameter set. Since the parent sample is volume-limited and confined to a narrow redshift interval, the superthin and control samples are closely matched in redshift by construction, and no explicit redshift control is therefore applied. A detailed description of the control-sample construction method is provided in \appref{app:const_contr}.

\subsection{Projected cross-correlation function}
\label{sec:rel_env_sup}

We first examine the projected cross-correlation function, $w_p(r_p)$, measured by cross-correlating each study sample with the SDSS reference sample described in Section~2.3 and projecting the two-dimensional correlation function along the line of sight, following \citet{li2006a, li2006b}. This statistic traces the average excess number of neighbouring reference galaxies around superthin or control galaxies as a function of projected separation.
For a given study sample, the cross pairs are counted with the reference sample ($QD$) and with the random sample ($QR$), respectively, within each bin of projected and line-of-sight separation $(r_p, \pi)$. The two-point cross-correlation function $\xi(r_p, \pi)$ is then estimated as
\begin{equation}
\xi(r_p, \pi)=\frac{N_R}{N_D}\frac{QD(r_p, \pi)}{QR(r_p, \pi)}-1,
\label{eq:corr}
\end{equation}
where $N_D$ and $N_R$ are the total numbers of galaxies in the reference and random samples, respectively. The projected cross-correlation function is obtained by integrating $\xi(r_p, \pi)$ along the line of sight:
\begin{equation}
w_p(r_p)=\int ^{+\infty}_{-\infty} \xi(r_p, \pi) d\pi =\sum_i \xi(r_p, \pi_i) \Delta \pi.
\label{eq:proj_corr}
\end{equation}
In practice, the summation is carried out over $-39.5 < \pi/(h^{-1}\mathrm{Mpc}) < 39.5$ in steps of $\Delta \pi = 1\,h^{-1}\mathrm{Mpc}$. The uncertainties are estimated by bootstrap resampling of the study sample. Further details can be found in \citet{li2006a, li2006b}.

\figref{fig:cross_2bins} shows that superthin galaxies have a lower clustering amplitude than their matched controls on projected scales of $\sim0.1$--$1\,h^{-1}\,\mathrm{Mpc}$. A similar scale-dependent anti-bias was found by \citet{li2006b} for narrow-line AGN, and their mock-catalogue analysis showed that it can be naturally explained if the study sample contains a higher fraction of central galaxies than the matched controls. By analogy, we interpret the intermediate-scale depression found here as a halo-occupation signal: the observed superthin sample may contain a higher central-galaxy fraction and a lower satellite fraction than the matched control sample. This inference is indirect for the SDSS sample, because the present data do not assign individual galaxies to central or satellite status. On scales of several $h^{-1}\,\mathrm{Mpc}$ and above, the superthin and control samples have similar clustering amplitudes, implying similar large-scale bias and hence comparable typical halo masses and large-scale environments for the matched samples. Since the galaxies in \figref{fig:color_mass} lie almost entirely below $\log_{10}(M_\ast/h^{-2} M_\odot)\sim10$, their host halos are still expected to be modest in absolute mass, but the relative superthin-control difference in $w_p(r_p)$ mainly reflects the central/satellite mix rather than a lower halo mass for superthin galaxies relative to their controls. The bulgeless and bulge subsamples show the same qualitative behaviour, and we find no clear evidence that the environmental trend depends on the presence of a fitted bulge.

 \begin{figure}
    \includegraphics[width=\columnwidth]{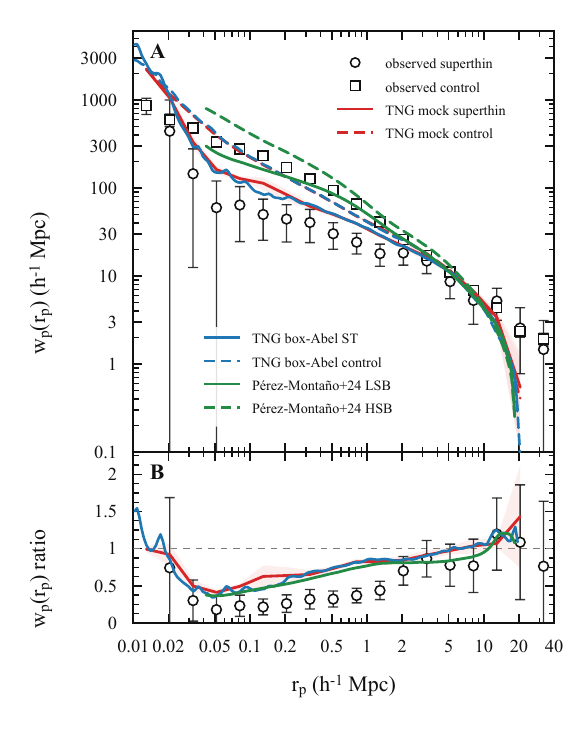}
    \caption{Comparison of the observed clustering of superthin galaxies with related measurements from IllustrisTNG and from simulated low surface brightness galaxies. The upper panel shows $w_p(r_p)$, and the lower panel shows the ratio relative to the corresponding control or high surface brightness sample. Black open symbols show the observed superthin and control samples. Red curves show mock-lightcone measurements for the TNG superthin and control samples, blue curves show the Abel projection of their real-space TNG correlation functions, and green curves show the rotation-dominated LSB and HSB measurements of \citet{perezmontano2024}, Abel-projected from their Fig.~7.}
    \label{fig:lsb_tng_wp_comparison}
\end{figure}

\figref{fig:lsb_tng_wp_comparison} further compares the observed result with superthin galaxies selected in IllustrisTNG by \citet{hu2024}. We show two simulation measurements: mock-lightcone projected correlations using a flux-limited reference sample, and Abel projections of the real-space TNG correlation functions. The observed and simulated superthin/control comparisons are qualitatively consistent: both show suppressed relative clustering on $\sim0.1$--$1\,h^{-1}\,\mathrm{Mpc}$ scales, as expected if superthin galaxies have a higher central-galaxy fraction and lower satellite fraction than their controls. In both observations and simulations, the ratios approach unity on larger scales, consistent with similar large-scale bias and comparable typical halo masses for each matched superthin-control pair. Quantitatively, however, the difference between superthin and control galaxies is milder in the simulation, so the simulated superthin/control ratios are closer to unity than the observed ratio over much of the $\sim0.1$--$1\,h^{-1}\,\mathrm{Mpc}$ range.

The central--satellite classifications available in IllustrisTNG provide a direct check of this interpretation for the simulated analogues. For the TNG catalogues used here, the central fraction is $312/380=0.82$ for the superthin sample, compared with $2186/3148=0.69$ for the control sample; a two-proportion test gives $p\simeq2.9\times10^{-7}$. Thus the TNG superthin galaxies do have a higher central fraction than their controls, supporting the interpretation that the suppressed intermediate-scale clustering in the simulation is mainly a halo-occupation effect. This result also gives indirect support to applying the same interpretation to the observed clustering trend. A complementary check using the TNG large-scale-environment grid gives the same qualitative picture: the simulated superthin galaxies have slightly lower $1\,\mathrm{Mpc}$ overdensities and a lower cluster/knot fraction than their controls, with much of this difference associated with their enhanced central fraction. The smaller central-fraction contrast in the simulation is also consistent with the simulated superthin/control clustering ratio being closer to unity than the observed one.

We also compare with low surface brightness galaxies, since superthin galaxies are often interpreted as their edge-on counterparts. \citet{perezmontano2024}
measured the real-space two-point correlation functions of rotation-dominated LSBGs and HSBGs in IllustrisTNG; for the comparison in \figref{fig:lsb_tng_wp_comparison}, we Abel-transform their real-space correlation functions into projected $w_p(r_p)$. The LSB/HSB ratio is broadly similar to the superthin/control ratios over intermediate scales, suggesting that a similar halo-occupation effect may also be present for diffuse, rotation-supported disk systems. At the same time, the absolute clustering amplitudes of both the LSB and HSB samples are systematically higher than those of the observed and simulated superthin/control samples. This difference is important and probably reflects, at least in part, the different selections: superthin galaxies are likely a rare and extreme subset of the broader LSBG population, selected by very large apparent disk thinness, whereas LSBG samples selected by surface brightness include a wider range of intrinsic shapes, inclinations, and halo environments. The comparison therefore supports a physical connection between superthin and LSB galaxies, but it should not be interpreted as a one-to-one quantitative equivalence between the two populations.

 \subsection{Local overdensity}

\begin{figure*}
    \includegraphics[width=\textwidth]{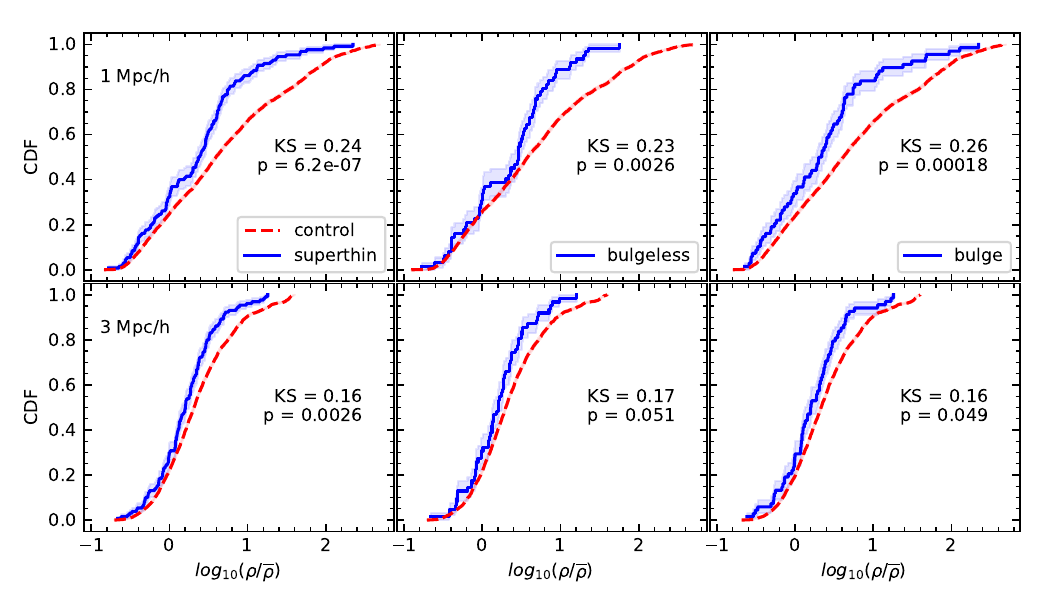}

    \caption{Cumulative distributions of local overdensity, shown as a function of $\log_{10}(\rho/\bar{\rho})$, for the superthin samples and their matched control samples. Line styles and panel columns are the same as in \figref{fig:cross_2bins}. The upper and lower rows correspond to smoothing scales of 1 and $3\,h^{-1}\,\mathrm{Mpc}$, respectively. The shaded regions represent the binomial uncertainty, $\sqrt{p(1-p)/N}$. The Kolmogorov--Smirnov (KS) test statistic and its p-value (testing the null hypothesis of identical distributions) are shown on the right side of each panel.}
    \label{fig:local_den_2bins}
\end{figure*}

We next use the reconstructed density field from \citet{Wang2016ELUCID} to compare the local overdensities of superthin and control galaxies. We consider smoothing scales of 1, 2, and $3\,h^{-1}\,\mathrm{Mpc}$; for compactness, \figref{fig:local_den_2bins} shows the results for 1 and $3\,h^{-1}\,\mathrm{Mpc}$, while the $2\,h^{-1}\,\mathrm{Mpc}$ result lies between them. Two-sample Kolmogorov--Smirnov (KS) test results (statistic and p-value) are shown on the right side of each panel.

The full superthin sample is shifted toward lower local densities relative to its control sample, especially at the $1\,h^{-1}\,\mathrm{Mpc}$ smoothing scale. The difference becomes weaker at $3\,h^{-1}\,\mathrm{Mpc}$, indicating that the environmental dependence is strongest on scales comparable to the sizes of dark matter halos; this weakening is also consistent with the similar large-scale clustering amplitudes seen in \figref{fig:cross_2bins}. The bulgeless and bulge subsamples show similar trends, with no clear evidence that the presence of a bulge changes the environmental preference. This local-density measurement is complementary to, but distinct from, the projected-correlation interpretation above. The latter mainly points to a higher central-galaxy fraction and lower satellite fraction, whereas the former directly shows lower reconstructed density on $1\,h^{-1}\,\mathrm{Mpc}$ scales. Together, they suggest that superthin galaxies are less often satellites in rich halo environments and, in our sample, occupy locally less crowded halo-scale regions.

Such a halo-scale preference is qualitatively in line with the formation scenario proposed by \citet{hu2024}, who found in the TNG100 simulation that superthin galaxies develop extended thin disks since $z\sim1$ as frequent prograde mergers increase the spin of their host dark matter halos. The wavelength-dependent disk-size trend found in \secref{sec:thin_band} is also compatible with a growth picture in which the disk appears larger in the younger stellar population traced more strongly by optical light. However, these links remain interpretive: our environmental and waveband measurements do not directly constrain halo spin or merger history. Direct measurements of halo spin proxies, central/satellite status, and galaxy kinematics would be needed to test this formation picture more directly.

\subsection{Large-scale structure}

Finally, we examine the LSS classification of the environments of superthin galaxies. Each galaxy is assigned to one of four cosmic-web types---void, sheet, filament, or cluster---as described in \secref{subsec:dens_lss}. Because the control sample used here is additionally matched in the $1\,h^{-1}\,\mathrm{Mpc}$ overdensity, this comparison tests for residual dependence on LSS type after controlling for both galaxy properties and smaller-scale environment.

\begin{figure*}
    \includegraphics[width=\textwidth]{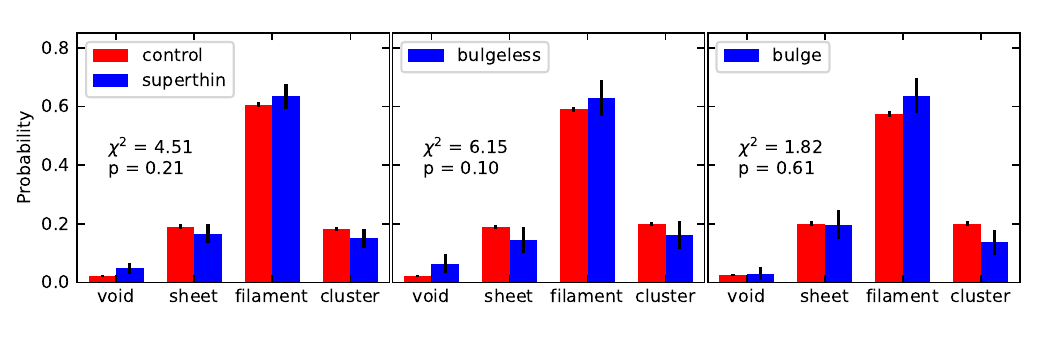}
    \caption{Fractions of galaxies in different large-scale-structure types for the superthin samples (blue) and their matched control samples (red). Panel columns are the same as in \figref{fig:cross_2bins}. The error bars represent the binomial uncertainty, $\sqrt{p(1-p)/N}$. The Pearson's chi-squared test statistic and its p-value (testing the null hypothesis of identical fractions) are shown in the left-centre region of each panel.}
    \label{fig:lss_2bins}
\end{figure*}

As shown in \figref{fig:lss_2bins}, the LSS fractions of superthin galaxies are consistent with those of the matched controls for the full sample and for both bulge subsamples, as further indicated by Pearson's chi-squared test results. Thus, after matching in stellar mass, colour, and local density, we find no clear residual dependence of superthin galaxies on cosmic-web type. This weak link with LSS is consistent with the results from the projected correlation function and local overdensity: the environmental signal is mainly confined to halo scales rather than to the larger-scale cosmic web. This conclusion differs from \citet{bizyaev2017}, who found that superthin galaxies are less connected to filaments. The discrepancy may be partly due to differences between their sample and ours, as already seen in the comparison of structural measurements in \secref{subsec:comp_bizyaev}. In addition, their filament-distance signal may partly reflect environmental differences on smaller scales, which are explicitly controlled for in our LSS comparison.

\section{Summary and Discussion}
\label{sec:summ}

In this work, we construct a sample of superthin galaxies selected from their disk shapes in the SDSS $r$ band, as measured through two-dimensional bulge/disk decomposition. We use SDSS $griz$ and UKIDSS $JHK$ imaging to test whether optically identified superthin galaxies remain superthin in the NIR, and we compare our measurements with previous 1D analyses of the B17 comparison sample and the B20 measurements. We also investigate the environments of the superthin galaxies using projected cross-correlation functions, reconstructed local overdensities, and LSS classifications, always comparing with control samples matched in the relevant galaxy properties. Our conclusions are summarised as follows:
\begin{enumerate}
    \item From the optical to the NIR bands, the disk scale sizes of superthin galaxies decrease along both the major and minor axes. The decreases along the two axes have roughly the same amplitude ($K/r \sim 0.6$), resulting in an almost invariant disk shape. The presence of a bulge has no significant influence on this result (Figures \ref{fig:demos_bands}, \ref{fig:comp_ba_bands}, \ref{fig:ba_bands});
    \item The apparent NIR thickening reported by B20 is mainly associated with the combination of sample selection and measurement method. In particular, the 1D method shows a magnitude- and band-dependent bias in the measured radial scale length, which can artificially increase $h_s/r_s$ in the NIR for fainter galaxies. When the 2D method is applied, both our sample and the B17 sample show little waveband dependence in disk thickness (Figures \ref{fig:comp_1d_2d}, \ref{fig:comp_rs_mag_r});
    \item Superthin galaxies show a halo-scale environmental difference. Their lower projected clustering amplitude on $\sim0.1$--$1\,h^{-1}\,\mathrm{Mpc}$ scales is best interpreted, following \citet{li2006b}, as a higher central-galaxy fraction and lower satellite fraction, while their similar clustering on scales of several $h^{-1}\,\mathrm{Mpc}$ and above indicates comparable large-scale bias and typical halo masses relative to the controls. Their lower overdensities at $1\,h^{-1}\,\mathrm{Mpc}$ provide a complementary local-density measurement. The presence of a bulge has no significant influence on this trend (Figures \ref{fig:cross_2bins}, \ref{fig:lsb_tng_wp_comparison}, \ref{fig:local_den_2bins});
	\item After controlling for stellar mass, colour, and local overdensity, superthin galaxies show no clear residual dependence on large-scale-structure type (\figref{fig:lss_2bins}). This indicates that their environmental connection is mainly a halo-scale effect rather than a direct dependence on the larger-scale cosmic web.
\end{enumerate}

Taken together, the structural and environmental results suggest that the extreme thinness of these galaxies is connected to their halo-scale environments and may be related to the angular-momentum assembly of their host halos, although this connection is not directly measured in the present work. The weak dependence of disk shape on waveband shows that we find no evidence for an NIR-dominant thick component prominent enough to alter the overall fitted disk shape. A thick disk component could still be present, but it is not dominant in the present single-disk decomposition and image depth. The increase in disk size from the NIR to the optical bands hints at disk growth from older to younger stellar populations, but this trend does not by itself identify the underlying growth mechanism.

Meanwhile, the inferred higher central-galaxy fraction, lower satellite contribution, and lower local overdensity on $1\,h^{-1}\,\mathrm{Mpc}$ scales suggest that superthin galaxies are less affected by environmental processes such as satellite interactions or tidal heating in massive halos, even though their large-scale clustering implies halo masses comparable to those of the matched controls. This picture is consistent with the formation scenario of \citet{hu2024}, in which superthin galaxies develop their extended thin disks since $z\sim1$ as frequent prograde mergers increase the spin of their host halos, and with the broader connection between specific angular momentum retention and disk morphology discussed by \citet{romeo2023}. In this interpretation, mergers are not simply absent; rather, their orbital configuration and their effect on halo angular momentum are crucial. This point also clarifies how our result relates to the argument of \citet{haslbauer2022}, who emphasized that the high observed abundance of thin disks is difficult to reproduce in current cosmological simulations. Our data do not by themselves resolve that broader population-level issue, but they suggest that, for the extreme superthin subset, the relevant question is not merely whether mergers occurred, but whether the merger history preserved or enhanced the angular momentum of the disk-forming system.

It is also important to acknowledge that alternative scenarios cannot be formally ruled out. For instance, a relatively quiescent merger history or a low level of internal heating would naturally allow disks to retain thin configurations over cosmic time. In the absence of direct constraints on central/satellite status, merger history, halo spin, or stellar/gas kinematics for individual galaxies, definitively distinguishing between these possibilities remains beyond the reach of the present data. Future work combining group catalogues, gas and stellar kinematics, and simulation-based mock observations will be important for testing this halo-spin interpretation more directly. Looking further ahead, existing and future deep imaging surveys such as HSC and Euclid, together with high-redshift spectroscopic surveys such as DESI and PFS, will enable systematic searches for superthin galaxies beyond the local universe, opening the possibility of tracing the evolution of this special class of galaxies over cosmic time.

\section*{Acknowledgements}

We are grateful to the anonymous referee for the helpful comments.
This work is supported by the National Science Foundation
of China (grant No. 12433003) and the National Key R\&D Program of China
(grant No. 2018YFA0404502). We thank Luis Enrique P\'erez-Monta\~no for his helpful comment and for providing the data from Fig.~7 of \citet{perezmontano2024}.

\section*{Data Availability}

This work uses datasets from sources in the public domain: the Sloan Digital Sky Survey (SDSS; \url{https://www.sdss.org/}), the UKIRT Infrared Deep Sky Survey (UKIDSS; \url{http://www.ukidss.org/}), the NASA Sloan Atlas (NSA; \url{http://www.nsatlas.org/}), and the New York University Value-Added Galaxy Catalog (NYU-VAGC; \url{http://sdss.physics.nyu.edu/vagc/}). The local density and large scale structure data were kindly provided by the authors of \citet{Wang2016ELUCID}. The newly generated data underlying this article are available upon reasonable request from
the authors.

\bibliography{references}

@INPROCEEDINGS{lupton2001,
       author = {{Lupton}, R. and {Gunn}, J.~E. and {Ivezi{\'c}}, Z. and {Knapp}, G.~R. and {Kent}, S.},
        title = "{The SDSS Imaging Pipelines}",
    booktitle = {Astronomical Data Analysis Software and Systems X},
         year = 2001,
       editor = {{Harnden}, Jr., F.~R. and {Primini}, Frances A. and {Payne}, Harry E.},
       series = {Astronomical Society of the Pacific Conference Series},
       volume = {238},
        month = jan,
    publisher = {ASP},
        pages = {269},
          doi = {10.48550/arXiv.astro-ph/0101420},
archivePrefix = {arXiv},
       eprint = {astro-ph/0101420},
 primaryClass = {astro-ph},
       adsurl = {https://ui.adsabs.harvard.edu/abs/2001ASPC..238..269L}
}

@ARTICLE{perezmontano2024,
       author = {{P{\'e}rez-Monta{\~n}o}, Luis Enrique and {Cervantes Sodi}, Bernardo and {Rodriguez-Gomez}, Vicente and {Zhu}, Qirong and {Ogiya}, Go},
        title = "{Environmental effects on low surface brightness galaxies in the IllustrisTNG simulation}",
      journal = {\mnras},
         year = 2024,
        month = sep,
       volume = {533},
       number = {1},
        pages = {93-108},
          doi = {10.1093/mnras/stae1793},
archivePrefix = {arXiv},
       eprint = {2407.15529},
 primaryClass = {astro-ph.GA},
       adsurl = {https://ui.adsabs.harvard.edu/abs/2024MNRAS.533...93P}
}

@ARTICLE{romeo2023,
       author = {{Romeo}, Alessandro B. and {Agertz}, Oscar and {Renaud}, Florent},
        title = "{The specific angular momentum of disc galaxies and its connection with galaxy morphology, bar structure, and disc gravitational instability}",
      journal = {\mnras},
         year = 2023,
        month = jan,
       volume = {518},
       number = {1},
        pages = {1002-1021},
          doi = {10.1093/mnras/stac3074},
archivePrefix = {arXiv},
       eprint = {2204.02695},
 primaryClass = {astro-ph.GA},
       adsurl = {https://ui.adsabs.harvard.edu/abs/2023MNRAS.518.1002R}
}

@ARTICLE{haslbauer2022,
       author = {{Haslbauer}, Moritz and {Banik}, Indranil and {Kroupa}, Pavel and {Wittenburg}, Nils and {Javanmardi}, Behnam},
        title = "{The High Fraction of Thin Disk Galaxies Continues to Challenge {\ensuremath{\Lambda}}CDM Cosmology}",
      journal = {\apj},
         year = 2022,
        month = feb,
       volume = {925},
       number = {2},
          eid = {183},
        pages = {183},
          doi = {10.3847/1538-4357/ac46ac},
archivePrefix = {arXiv},
       eprint = {2202.01221},
 primaryClass = {astro-ph.GA},
       adsurl = {https://ui.adsabs.harvard.edu/abs/2022ApJ...925..183H}
}

@INBOOK{voron1967,
       author = {{Vorontsov-Velyaminov}, B.},
        title = "{New morphological types of galaxies}",
    booktitle = {Modern astrophysics. A memorial to Otto Struve},
         year = 1967,
    publisher = {Gauthier-Villars, Paris},
        pages = {347},
       adsurl = {https://ui.adsabs.harvard.edu/abs/1967mamt.book..347V}
}

@ARTICLE{devau1974,
       author = {{de Vaucouleurs}, G.},
        title = "{Structure, Dynamics and Statistical Properties of Galaxies (invited Paper)}",
      journal = {Symposium - International Astronomical Union},
         year = 1974,
       volume = {58},
        month = jan,
        pages = {1-53},
          doi = {10.1017/S007418090002430X},
       adsurl = {https://ui.adsabs.harvard.edu/abs/1974IAUS...58....1D}
}

@ARTICLE{Wang2016ELUCID,
       author = {{Wang}, Huiyuan and {Mo}, H.~J. and {Yang}, Xiaohu and {Zhang}, Youcai and {Shi}, JingJing and {Jing}, Y.~P. and {Liu}, Chengze and {Li}, Shijie and {Kang}, Xi and {Gao}, Yang},
        title = "{ELUCID - Exploring the Local Universe with ReConstructed Initial Density Field III: Constrained Simulation in the SDSS Volume}",
      journal = {\apj},
         year = 2016,
        month = nov,
       volume = {831},
       number = {2},
          eid = {164},
        pages = {164},
          doi = {10.3847/0004-637X/831/2/164},
archivePrefix = {arXiv},
       eprint = {1608.01763},
 primaryClass = {astro-ph.CO},
       adsurl = {https://ui.adsabs.harvard.edu/abs/2016ApJ...831..164W}
}

@ARTICLE{SDSS-DR7,
       author = {{Abazajian}, Kevork N. and {Adelman-McCarthy}, Jennifer K. and {Ag{\"u}eros}, Marcel A. and {Allam}, Sahar S. and {Allende Prieto}, Carlos and {An}, Deokkeun and {Anderson}, Kurt S.~J. and {Anderson}, Scott F. and {Annis}, James and {Bahcall}, Neta A. and {Bailer-Jones}, C.~A.~L. and {Barentine}, J.~C. and {Bassett}, Bruce A. and {Becker}, Andrew C. and {Beers}, Timothy C. and {Bell}, Eric F. and {Belokurov}, Vasily and {Berlind}, Andreas A. and {Berman}, Eileen F. and {Bernardi}, Mariangela and {Bickerton}, Steven J. and {Bizyaev}, Dmitry and {Blakeslee}, John P. and {Blanton}, Michael R. and {Bochanski}, John J. and {Boroski}, William N. and {Brewington}, Howard J. and {Brinchmann}, Jarle and {Brinkmann}, J. and {Brunner}, Robert J. and {Budav{\'a}ri}, Tam{\'a}s and {Carey}, Larry N. and {Carliles}, Samuel and {Carr}, Michael A. and {Castander}, Francisco J. and {Cinabro}, David and {Connolly}, A.~J. and {Csabai}, Istv{\'a}n and {Cunha}, Carlos E. and {Czarapata}, Paul C. and {Davenport}, James R.~A. and {de Haas}, Ernst and {Dilday}, Ben and {Doi}, Mamoru and {Eisenstein}, Daniel J. and {Evans}, Michael L. and {Evans}, N.~W. and {Fan}, Xiaohui and {Friedman}, Scott D. and {Frieman}, Joshua A. and {Fukugita}, Masataka and {G{\"a}nsicke}, Boris T. and {Gates}, Evalyn and {Gillespie}, Bruce and {Gilmore}, G. and {Gonzalez}, Belinda and {Gonzalez}, Carlos F. and {Grebel}, Eva K. and {Gunn}, James E. and {Gy{\"o}ry}, Zsuzsanna and {Hall}, Patrick B. and {Harding}, Paul and {Harris}, Frederick H. and {Harvanek}, Michael and {Hawley}, Suzanne L. and {Hayes}, Jeffrey J.~E. and {Heckman}, Timothy M. and {Hendry}, John S. and {Hennessy}, Gregory S. and {Hindsley}, Robert B. and {Hoblitt}, J. and {Hogan}, Craig J. and {Hogg}, David W. and {Holtzman}, Jon A. and {Hyde}, Joseph B. and {Ichikawa}, Shin-ichi and {Ichikawa}, Takashi and {Im}, Myungshin and {Ivezi{\'c}}, {\v{Z}}eljko and {Jester}, Sebastian and {Jiang}, Linhua and {Johnson}, Jennifer A. and {Jorgensen}, Anders M. and {Juri{\'c}}, Mario and {Kent}, Stephen M. and {Kessler}, R. and {Kleinman}, S.~J. and {Knapp}, G.~R. and {Konishi}, Kohki and {Kron}, Richard G. and {Krzesinski}, Jurek and {Kuropatkin}, Nikolay and {Lampeitl}, Hubert and {Lebedeva}, Svetlana and {Lee}, Myung Gyoon and {Lee}, Young Sun and {French Leger}, R. and {L{\'e}pine}, S{\'e}bastien and {Li}, Nolan and {Lima}, Marcos and {Lin}, Huan and {Long}, Daniel C. and {Loomis}, Craig P. and {Loveday}, Jon and {Lupton}, Robert H. and {Magnier}, Eugene and {Malanushenko}, Olena and {Malanushenko}, Viktor and {Mandelbaum}, Rachel and {Margon}, Bruce and {Marriner}, John P. and {Mart{\'\i}nez-Delgado}, David and {Matsubara}, Takahiko and {McGehee}, Peregrine M. and {McKay}, Timothy A. and {Meiksin}, Avery and {Morrison}, Heather L. and {Mullally}, Fergal and {Munn}, Jeffrey A. and {Murphy}, Tara and {Nash}, Thomas and {Nebot}, Ada and {Neilsen}, Eric H., Jr. and {Newberg}, Heidi Jo and {Newman}, Peter R. and {Nichol}, Robert C. and {Nicinski}, Tom and {Nieto-Santisteban}, Maria and {Nitta}, Atsuko and {Okamura}, Sadanori and {Oravetz}, Daniel J. and {Ostriker}, Jeremiah P. and {Owen}, Russell and {Padmanabhan}, Nikhil and {Pan}, Kaike and {Park}, Changbom and {Pauls}, George and {Peoples}, John, Jr. and {Percival}, Will J. and {Pier}, Jeffrey R. and {Pope}, Adrian C. and {Pourbaix}, Dimitri and {Price}, Paul A. and {Purger}, Norbert and {Quinn}, Thomas and {Raddick}, M. Jordan and {Re Fiorentin}, Paola and {Richards}, Gordon T. and {Richmond}, Michael W. and {Riess}, Adam G. and {Rix}, Hans-Walter and {Rockosi}, Constance M. and {Sako}, Masao and {Schlegel}, David J. and {Schneider}, Donald P. and {Scholz}, Ralf-Dieter and {Schreiber}, Matthias R. and {Schwope}, Axel D. and {Seljak}, Uro{\v{s}} and {Sesar}, Branimir and {Sheldon}, Erin and {Shimasaku}, Kazu and {Sibley}, Valena C. and {Simmons}, A.~E. and {Sivarani}, Thirupathi and {Allyn Smith}, J. and {Smith}, Martin C. and {Smol{\v{c}}i{\'c}}, Vernesa and {Snedden}, Stephanie A. and {Stebbins}, Albert and {Steinmetz}, Matthias and {Stoughton}, Chris and {Strauss}, Michael A. and {SubbaRao}, Mark and {Suto}, Yasushi and {Szalay}, Alexander S. and {Szapudi}, Istv{\'a}n and {Szkody}, Paula and {Tanaka}, Masayuki and {Tegmark}, Max and {Teodoro}, Luis F.~A. and {Thakar}, Aniruddha R. and {Tremonti}, Christy A. and {Tucker}, Douglas L. and {Uomoto}, Alan and {Vanden Berk}, Daniel E. and {Vandenberg}, Jan and {Vidrih}, S. and {Vogeley}, Michael S. and {Voges}, Wolfgang and {Vogt}, Nicole P. and {Wadadekar}, Yogesh and {Watters}, Shannon and {Weinberg}, David H. and {West}, Andrew A. and {White}, Simon D.~M. and {Wilhite}, Brian C. and {Wonders}, Alainna C. and {Yanny}, Brian and {Yocum}, D.~R. and {York}, Donald G. and {Zehavi}, Idit and {Zibetti}, Stefano and {Zucker}, Daniel B.},
        title = "{The Seventh Data Release of the Sloan Digital Sky Survey}",
      journal = {\apjs},
         year = 2009,
        month = jun,
       volume = {182},
       number = {2},
        pages = {543-558},
          doi = {10.1088/0067-0049/182/2/543},
archivePrefix = {arXiv},
       eprint = {0812.0649},
 primaryClass = {astro-ph},
       adsurl = {https://ui.adsabs.harvard.edu/abs/2009ApJS..182..543A}
}

@ARTICLE{aditya2025,
       author = {{Aditya}, K. and {Banerjee}, Arunima},
        title = "{Evolution of Low Surface Brightness Ultrathin Galaxies: The Role of Dark Matter Halo and Bar Formation on Disk Thickness}",
      journal = {\apj},
         year = 2025,
        month = oct,
       volume = {992},
       number = {2},
          eid = {179},
        pages = {179},
          doi = {10.3847/1538-4357/ae0991},
archivePrefix = {arXiv},
       eprint = {2509.12966},
 primaryClass = {astro-ph.GA},
       adsurl = {https://ui.adsabs.harvard.edu/abs/2025ApJ...992..179A}
}

@ARTICLE{marinacci2018,
       author = {{Marinacci}, Federico and {Vogelsberger}, Mark and {Pakmor}, R{\"u}diger and {Torrey}, Paul and {Springel}, Volker and {Hernquist}, Lars and {Nelson}, Dylan and {Weinberger}, Rainer and {Pillepich}, Annalisa and {Naiman}, Jill and {Genel}, Shy},
        title = "{First results from the IllustrisTNG simulations: radio haloes and magnetic fields}",
      journal = {\mnras},
         year = 2018,
        month = nov,
       volume = {480},
       number = {4},
        pages = {5113-5139},
          doi = {10.1093/mnras/sty2206},
archivePrefix = {arXiv},
       eprint = {1707.03396},
 primaryClass = {astro-ph.CO},
       adsurl = {https://ui.adsabs.harvard.edu/abs/2018MNRAS.480.5113M}
}

@ARTICLE{naiman2018,
       author = {{Naiman}, Jill P. and {Pillepich}, Annalisa and {Springel}, Volker and {Ramirez-Ruiz}, Enrico and {Torrey}, Paul and {Vogelsberger}, Mark and {Pakmor}, R{\"u}diger and {Nelson}, Dylan and {Marinacci}, Federico and {Hernquist}, Lars and {Weinberger}, Rainer and {Genel}, Shy},
        title = "{First results from the IllustrisTNG simulations: a tale of two elements - chemical evolution of magnesium and europium}",
      journal = {\mnras},
         year = 2018,
        month = jun,
       volume = {477},
       number = {1},
        pages = {1206-1224},
          doi = {10.1093/mnras/sty618},
archivePrefix = {arXiv},
       eprint = {1707.03401},
 primaryClass = {astro-ph.GA},
       adsurl = {https://ui.adsabs.harvard.edu/abs/2018MNRAS.477.1206N}
}

@ARTICLE{nelson2018,
       author = {{Nelson}, Dylan and {Pillepich}, Annalisa and {Springel}, Volker and {Weinberger}, Rainer and {Hernquist}, Lars and {Pakmor}, R{\"u}diger and {Genel}, Shy and {Torrey}, Paul and {Vogelsberger}, Mark and {Kauffmann}, Guinevere and {Marinacci}, Federico and {Naiman}, Jill},
        title = "{First results from the IllustrisTNG simulations: the galaxy colour bimodality}",
      journal = {\mnras},
         year = 2018,
        month = mar,
       volume = {475},
       number = {1},
        pages = {624-647},
          doi = {10.1093/mnras/stx3040},
archivePrefix = {arXiv},
       eprint = {1707.03395},
 primaryClass = {astro-ph.GA},
       adsurl = {https://ui.adsabs.harvard.edu/abs/2018MNRAS.475..624N}
}

@ARTICLE{pillepich2018,
       author = {{Pillepich}, Annalisa and {Nelson}, Dylan and {Hernquist}, Lars and {Springel}, Volker and {Pakmor}, R{\"u}diger and {Torrey}, Paul and {Weinberger}, Rainer and {Genel}, Shy and {Naiman}, Jill P. and {Marinacci}, Federico and {Vogelsberger}, Mark},
        title = "{First results from the IllustrisTNG simulations: the stellar mass content of groups and clusters of galaxies}",
      journal = {\mnras},
         year = 2018,
        month = mar,
       volume = {475},
       number = {1},
        pages = {648-675},
          doi = {10.1093/mnras/stx3112},
archivePrefix = {arXiv},
       eprint = {1707.03406},
 primaryClass = {astro-ph.GA},
       adsurl = {https://ui.adsabs.harvard.edu/abs/2018MNRAS.475..648P}
}

@ARTICLE{springel2018,
       author = {{Springel}, Volker and {Pakmor}, R{\"u}diger and {Pillepich}, Annalisa and {Weinberger}, Rainer and {Nelson}, Dylan and {Hernquist}, Lars and {Vogelsberger}, Mark and {Genel}, Shy and {Torrey}, Paul and {Marinacci}, Federico and {Naiman}, Jill},
        title = "{First results from the IllustrisTNG simulations: matter and galaxy clustering}",
      journal = {\mnras},
         year = 2018,
        month = mar,
       volume = {475},
       number = {1},
        pages = {676-698},
          doi = {10.1093/mnras/stx3304},
archivePrefix = {arXiv},
       eprint = {1707.03397},
 primaryClass = {astro-ph.GA},
       adsurl = {https://ui.adsabs.harvard.edu/abs/2018MNRAS.475..676S}
}

@ARTICLE{hu2024,
       author = {{Hu}, Jianhong and {Xu}, Dandan and {Li}, Cheng},
        title = "{Formation of Superthin Galaxies in IllustrisTNG}",
      journal = {Research in Astronomy and Astrophysics},
         year = 2024,
        month = jul,
       volume = {24},
       number = {7},
          eid = {075019},
        pages = {075019},
          doi = {10.1088/1674-4527/ad5399},
archivePrefix = {arXiv},
       eprint = {2406.13745},
 primaryClass = {astro-ph.GA},
       adsurl = {https://ui.adsabs.harvard.edu/abs/2024RAA....24g5019H}
}

@article{banerjee2017,
    author = {{Banerjee}, A. and {Bapat}, D.},
    title = "{Mass modelling of superthin galaxies: IC5249, UGC7321 and IC2233}",
    journal = {\mnras},
    year = 2017,
    month = apr,
    volume = 466,
    number = 3,
    pages = {3753-3761},
    doi = {10.1093/mnras/stw3219},
    adsurl = {https://ui.adsabs.harvard.edu/abs/2017MNRAS.466.3753B}
}

@article{narayanan2022,
    author = {{Narayanan}, G. and {Banerjee}, A.},
    title = "{Are superthin galaxies low-surface-brightness galaxies seen edge-on? The star formation probe}",
    journal = {\mnras},
    year = 2022,
    month = aug,
    volume = 514,
    number = 4,
    pages = {5126-5140},
    doi = {10.1093/mnras/stac1662},
    adsurl = {https://ui.adsabs.harvard.edu/abs/2022MNRAS.514.5126N}
}

@ARTICLE{peng2010,
       author = {{Peng}, Chien Y. and {Ho}, Luis C. and {Impey}, Chris D. and {Rix}, Hans-Walter},
        title = "{Detailed Decomposition of Galaxy Images. II. Beyond Axisymmetric Models}",
      journal = {\aj},
         year = 2010,
        month = jun,
       volume = {139},
       number = {6},
        pages = {2097-2129},
          doi = {10.1088/0004-6256/139/6/2097},
archivePrefix = {arXiv},
       eprint = {0912.0731},
 primaryClass = {astro-ph.CO},
       adsurl = {https://ui.adsabs.harvard.edu/abs/2010AJ....139.2097P}
}

@ARTICLE{peng2002,
       author = {{Peng}, Chien Y. and {Ho}, Luis C. and {Impey}, Chris D. and {Rix}, Hans-Walter},
        title = "{Detailed Structural Decomposition of Galaxy Images}",
      journal = {\aj},
         year = 2002,
        month = jul,
       volume = {124},
       number = {1},
        pages = {266-293},
          doi = {10.1086/340952},
archivePrefix = {arXiv},
       eprint = {astro-ph/0204182},
 primaryClass = {astro-ph},
       adsurl = {https://ui.adsabs.harvard.edu/abs/2002AJ....124..266P}
}

@ARTICLE{blanton2011,
       author = {{Blanton}, Michael R. and {Kazin}, Eyal and {Muna}, Demitri and {Weaver}, Benjamin A. and {Price-Whelan}, Adrian},
        title = "{Improved Background Subtraction for the Sloan Digital Sky Survey Images}",
      journal = {\aj},
         year = 2011,
        month = jul,
       volume = {142},
       number = {1},
          eid = {31},
        pages = {31},
          doi = {10.1088/0004-6256/142/1/31},
archivePrefix = {arXiv},
       eprint = {1105.1960},
 primaryClass = {astro-ph.IM},
       adsurl = {https://ui.adsabs.harvard.edu/abs/2011AJ....142...31B}
}

@ARTICLE{bertin1996,
       author = {{Bertin}, E. and {Arnouts}, S.},
        title = "{SExtractor: Software for source extraction.}",
      journal = {\aaps},
         year = 1996,
        month = jun,
       volume = {117},
        pages = {393-404},
          doi = {10.1051/aas:1996164},
       adsurl = {https://ui.adsabs.harvard.edu/abs/1996A&AS..117..393B}
}

@ARTICLE{matthews1999,
       author = {{Matthews}, L.~D. and {Gallagher}, J.~S., III and {van Driel}, W.},
        title = "{The Extraordinary ``Superthin'' Spiral Galaxy UGC 7321. I. Disk Color Gradients and Global Properties from Multiwavelength Observations}",
      journal = {\aj},
         year = 1999,
        month = dec,
       volume = {118},
       number = {6},
        pages = {2751-2766},
          doi = {10.1086/301128},
archivePrefix = {arXiv},
       eprint = {astro-ph/9909142},
 primaryClass = {astro-ph},
       adsurl = {https://ui.adsabs.harvard.edu/abs/1999AJ....118.2751M}
}

@ARTICLE{li2006a,
       author = {{Li}, Cheng and {Kauffmann}, Guinevere and {Jing}, Y.~P. and {White}, Simon D.~M. and {B{\"o}rner}, Gerhard and {Cheng}, F.~Z.},
        title = "{The dependence of clustering on galaxy properties}",
      journal = {\mnras},
         year = 2006,
        month = may,
       volume = {368},
       number = {1},
        pages = {21-36},
          doi = {10.1111/j.1365-2966.2006.10066.x},
archivePrefix = {arXiv},
       eprint = {astro-ph/0509873},
 primaryClass = {astro-ph},
       adsurl = {https://ui.adsabs.harvard.edu/abs/2006MNRAS.368...21L}
}

@ARTICLE{li2006b,
       author = {{Li}, Cheng and {Kauffmann}, Guinevere and {Wang}, Lan and {White}, Simon D.~M. and {Heckman}, Timothy M. and {Jing}, Y.~P.},
        title = "{The clustering of narrow-line AGN in the local Universe}",
      journal = {\mnras},
         year = 2006,
        month = dec,
       volume = {373},
       number = {2},
        pages = {457-468},
          doi = {10.1111/j.1365-2966.2006.11079.x},
archivePrefix = {arXiv},
       eprint = {astro-ph/0607492},
 primaryClass = {astro-ph},
       adsurl = {https://ui.adsabs.harvard.edu/abs/2006MNRAS.373..457L}
}

@INPROCEEDINGS{goad1979,
       author = {{Goad}, J.~W. and {Roberts}, M.~S.},
        title = "{Spectroscopic Observations of Superthin Galaxies}",
    booktitle = {Bulletin of the American Astronomical Society},
         year = 1979,
       volume = {11},
        month = dec,
        pages = {668},
       adsurl = {https://ui.adsabs.harvard.edu/abs/1979BAAS...11..668G}
}

@ARTICLE{goad1981,
       author = {{Goad}, J.~W. and {Roberts}, M.~S.},
        title = "{Spectroscopic observations of superthin galaxies.}",
      journal = {\apj},
         year = 1981,
        month = nov,
       volume = {250},
        pages = {79-86},
          doi = {10.1086/159349},
       adsurl = {https://ui.adsabs.harvard.edu/abs/1981ApJ...250...79G}
}

@ARTICLE{karachentsev1989,
       author = {{Karachentsev}, Igor'},
        title = "{Thin Edge-On Galaxies as a Tool for the Investigation of Large-Scale Streaming Motions in the Universe}",
      journal = {\aj},
         year = 1989,
        month = jun,
       volume = {97},
        pages = {1566},
          doi = {10.1086/115098},
       adsurl = {https://ui.adsabs.harvard.edu/abs/1989AJ.....97.1566K}
}

@ARTICLE{karachentsev1993,
       author = {{Karachentsev}, I.~D. and {Karachentseva}, V.~E. and {Parnovskij}, S.~L.},
        title = "{Flat galaxies catalogue.}",
      journal = {Astronomische Nachrichten},
         year = 1993,
        month = may,
       volume = {314},
       number = {3},
        pages = {97-222},
          doi = {10.1002/asna.2113140302},
       adsurl = {https://ui.adsabs.harvard.edu/abs/1993AN....314...97K}
}

@ARTICLE{karachentsev1999,
       author = {{Karachentsev}, I.~D. and {Karachentseva}, V.~E. and {Kudrya}, Yu. N. and {Sharina}, M.~E. and {Parnovskij}, S.~L.},
        title = "{The revised Flat Galaxy Catalogue.}",
      journal = {Bulletin of the Special Astrophysics Observatory},
         year = 1999,
        month = jan,
       volume = {47},
        pages = {5-185},
archivePrefix = {arXiv},
       eprint = {astro-ph/0305566},
 primaryClass = {astro-ph},
       adsurl = {https://ui.adsabs.harvard.edu/abs/1999BSAO...47....5K}
}

@ARTICLE{kautsch2006,
       author = {{Kautsch}, S.~J. and {Grebel}, E.~K. and {Barazza}, F.~D. and {Gallagher}, J.~S., III},
        title = "{A catalog of edge-on disk galaxies. From galaxies with a bulge to superthin galaxies}",
      journal = {\aap},
         year = 2006,
        month = jan,
       volume = {445},
       number = {2},
        pages = {765-778},
          doi = {10.1051/0004-6361:20053981},
archivePrefix = {arXiv},
       eprint = {astro-ph/0509294},
 primaryClass = {astro-ph},
       adsurl = {https://ui.adsabs.harvard.edu/abs/2006A&A...445..765K}
}

@ARTICLE{kautsch2009a,
       author = {{Kautsch}, S.~J.},
        title = "{Edge-on disk galaxies in the SDSS DR6: Fractions of bulgeless and other disk galaxies}",
      journal = {Astronomische Nachrichten},
         year = 2009,
        month = jan,
       volume = {330},
       number = {1},
        pages = {100},
          doi = {10.1002/asna.200811132},
archivePrefix = {arXiv},
       eprint = {0901.1894},
 primaryClass = {astro-ph.GA},
       adsurl = {https://ui.adsabs.harvard.edu/abs/2009AN....330..100K}
}

@ARTICLE{kautsch2009b,
       author = {{Kautsch}, S.~J.},
        title = "{Flat galaxies in the SDSS DR6}",
      journal = {Astronomische Nachrichten},
         year = 2009,
        month = dec,
       volume = {330},
        pages = {1053},
          doi = {10.1002/asna.200911293},
       adsurl = {https://ui.adsabs.harvard.edu/abs/2009AN....330.1053K}
}

@ARTICLE{kautsch2009c,
       author = {{Kautsch}, S.~J. and {Gallagher}, J.~S. and {Grebel}, E.~K.},
        title = "{Disk galaxies and their environment}",
      journal = {Astronomische Nachrichten},
         year = 2009,
        month = dec,
       volume = {330},
        pages = {1056},
          doi = {10.1002/asna.200911292},
       adsurl = {https://ui.adsabs.harvard.edu/abs/2009AN....330.1056K}
}

@ARTICLE{kautsch2009d,
       author = {{Kautsch}, Stefan J.},
        title = "{The Edge-On Perspective of Bulgeless, Simple Disk Galaxies}",
      journal = {\pasp},
         year = 2009,
        month = dec,
       volume = {121},
       number = {886},
        pages = {1297},
          doi = {10.1086/649607},
archivePrefix = {arXiv},
       eprint = {1001.4542},
 primaryClass = {astro-ph.CO},
       adsurl = {https://ui.adsabs.harvard.edu/abs/2009PASP..121.1297K}
}

@ARTICLE{bizyaev2002,
       author = {{Bizyaev}, D. and {Mitronova}, S.},
        title = "{Photometric parameters of edge-on galaxies from 2MASS observations}",
      journal = {\aap},
         year = 2002,
        month = jul,
       volume = {389},
        pages = {795-801},
          doi = {10.1051/0004-6361:20020633},
archivePrefix = {arXiv},
       eprint = {astro-ph/0207539},
 primaryClass = {astro-ph},
       adsurl = {https://ui.adsabs.harvard.edu/abs/2002A&A...389..795B}
}

@ARTICLE{bizyaev2009,
       author = {{Bizyaev}, Dmitry and {Mitronova}, Sofia},
        title = "{Structural Parameters of Stellar Disks from two Micron All Sky Survey Images of Edge-on Galaxies}",
      journal = {\apj},
         year = 2009,
        month = sep,
       volume = {702},
       number = {2},
        pages = {1567-1574},
          doi = {10.1088/0004-637X/702/2/1567},
archivePrefix = {arXiv},
       eprint = {0907.3472},
 primaryClass = {astro-ph.CO},
       adsurl = {https://ui.adsabs.harvard.edu/abs/2009ApJ...702.1567B}
}

@ARTICLE{bizyaev2014,
       author = {{Bizyaev}, D.~V. and {Kautsch}, S.~J. and {Mosenkov}, A.~V. and {Reshetnikov}, V.~P. and {Sotnikova}, N. Ya. and {Yablokova}, N.~V. and {Hillyer}, R.~W.},
        title = "{The Catalog of Edge-on Disk Galaxies from SDSS. I. The Catalog and the Structural Parameters of Stellar Disks}",
      journal = {\apj},
         year = 2014,
        month = may,
       volume = {787},
       number = {1},
          eid = {24},
        pages = {24},
          doi = {10.1088/0004-637X/787/1/24},
archivePrefix = {arXiv},
       eprint = {1404.3072},
 primaryClass = {astro-ph.GA},
       adsurl = {https://ui.adsabs.harvard.edu/abs/2014ApJ...787...24B}
}

@ARTICLE{bizyaev2017,
       author = {{Bizyaev}, D.~V. and {Kautsch}, S.~J. and {Sotnikova}, N. Ya. and {Reshetnikov}, V.~P. and {Mosenkov}, A.~V.},
        title = "{Very thin disc galaxies in the SDSS catalogue of edge-on galaxies}",
      journal = {\mnras},
         year = 2017,
        month = mar,
       volume = {465},
       number = {4},
        pages = {3784-3792},
          doi = {10.1093/mnras/stw2972},
archivePrefix = {arXiv},
       eprint = {1612.01042},
 primaryClass = {astro-ph.GA},
       adsurl = {https://ui.adsabs.harvard.edu/abs/2017MNRAS.465.3784B}
}

@ARTICLE{bizyaev2020,
       author = {{Bizyaev}, Dmitry and {Tatarnikov}, Andrey and {Shatsky}, Nikolai and {Najip}, Aurik and {Birlak}, Marina and {Voziakova}, Olga},
        title = "{Near-infrared photometry of superthin edge-on galaxies}",
      journal = {Astronomische Nachrichten},
         year = 2020,
        month = mar,
       volume = {341},
       number = {3},
        pages = {314-323},
          doi = {10.1002/asna.202013700},
archivePrefix = {arXiv},
       eprint = {2003.09842},
 primaryClass = {astro-ph.GA},
       adsurl = {https://ui.adsabs.harvard.edu/abs/2020AN....341..314B}
}

@ARTICLE{bizyaev2021,
       author = {{Bizyaev}, Dmitry and {Makarov}, D.~I. and {Reshetnikov}, V.~P. and {Mosenkov}, A.~V. and {Kautsch}, S.~J. and {Antipova}, A.~V.},
        title = "{Spectral Observations of Superthin Galaxies}",
      journal = {\apj},
         year = 2021,
        month = jun,
       volume = {914},
       number = {2},
          eid = {104},
        pages = {104},
          doi = {10.3847/1538-4357/abfb03},
archivePrefix = {arXiv},
       eprint = {2105.11855},
 primaryClass = {astro-ph.GA},
       adsurl = {https://ui.adsabs.harvard.edu/abs/2021ApJ...914..104B}
}

@ARTICLE{padilla2008,
       author = {{Padilla}, Nelson D. and {Strauss}, Michael A.},
        title = "{The shapes of galaxies in the Sloan Digital Sky Survey}",
      journal = {\mnras},
         year = 2008,
        month = aug,
       volume = {388},
       number = {3},
        pages = {1321-1334},
          doi = {10.1111/j.1365-2966.2008.13480.x},
archivePrefix = {arXiv},
       eprint = {0802.0877},
 primaryClass = {astro-ph},
       adsurl = {https://ui.adsabs.harvard.edu/abs/2008MNRAS.388.1321P}
}

@ARTICLE{holmberg1946,
       author = {{Holmberg}, Erik},
        title = "{On the Apparent Diameters and the Orientation in Space of Extragalactic Nebulae}",
      journal = {Meddelanden fran Lunds Astronomiska Observatorium Serie II},
         year = 1946,
        month = jan,
       volume = {117},
        pages = {3-82},
       adsurl = {https://ui.adsabs.harvard.edu/abs/1946MeLuS.117....3H}
}

@ARTICLE{sand1970,
       author = {{Sandage}, Allan and {Freeman}, Kenneth C. and {Stokes}, N.~R.},
        title = "{The Intrinsic Flattening of e, so, and Spiral Galaxies as Related to Galaxy Formation and Evolution}",
      journal = {\apj},
         year = 1970,
        month = jun,
       volume = {160},
        pages = {831},
          doi = {10.1086/150475},
       adsurl = {https://ui.adsabs.harvard.edu/abs/1970ApJ...160..831S}
}

@ARTICLE{noerdlinger1979,
       author = {{Noerdlinger}, P.~D.},
        title = "{The intrinsic flattening of galaxies.}",
      journal = {\apj},
         year = 1979,
        month = dec,
       volume = {234},
        pages = {802-809},
          doi = {10.1086/157559},
       adsurl = {https://ui.adsabs.harvard.edu/abs/1979ApJ...234..802N}
}

@ARTICLE{binney1981,
       author = {{Binney}, J. and {de Vaucouleurs}, G.},
        title = "{The apparent and true ellipticities of galaxies of different Hubble types in the Second Reference Catalogue.}",
      journal = {\mnras},
         year = 1981,
        month = feb,
       volume = {194},
        pages = {679-691},
          doi = {10.1093/mnras/194.3.679},
       adsurl = {https://ui.adsabs.harvard.edu/abs/1981MNRAS.194..679B}
}

@ARTICLE{heidmann1972,
       author = {{Heidmann}, Jean and {Heidmann}, Nicole and {de Vaucouleurs}, Gerard},
        title = "{Inclination and absorption effects on the apparent diameters, optical luminosities and neutral hydrogen radiation of galaxies{\textemdash}I. Optical and 21-cm line data}",
      journal = {\memras},
         year = 1972,
        month = jan,
       volume = {75},
        pages = {85},
       adsurl = {https://ui.adsabs.harvard.edu/abs/1972MmRAS..75...85H}
}

@ARTICLE{guthrie1992,
       author = {{Guthrie}, B.~N.~G.},
        title = "{Axial ratios of edge-on spirals.}",
      journal = {\aaps},
         year = 1992,
        month = may,
       volume = {93},
        pages = {255-270},
       adsurl = {https://ui.adsabs.harvard.edu/abs/1992A&AS...93..255G}
}

@ARTICLE{degrijs1998,
       author = {{de Grijs}, R.},
        title = "{The global structure of galactic discs}",
      journal = {\mnras},
         year = 1998,
        month = sep,
       volume = {299},
       number = {2},
        pages = {595-610},
          doi = {10.1046/j.1365-8711.1998.01896.x},
archivePrefix = {arXiv},
       eprint = {astro-ph/9804337},
 primaryClass = {astro-ph},
       adsurl = {https://ui.adsabs.harvard.edu/abs/1998MNRAS.299..595D}
}

@ARTICLE{kregel2002,
       author = {{Kregel}, Michiel and {van der Kruit}, Pieter C. and {de Grijs}, Richard},
        title = "{Flattening and truncation of stellar discs in edge-on spiral galaxies}",
      journal = {\mnras},
         year = 2002,
        month = aug,
       volume = {334},
       number = {3},
        pages = {646-668},
          doi = {10.1046/j.1365-8711.2002.05556.x},
archivePrefix = {arXiv},
       eprint = {astro-ph/0204154},
 primaryClass = {astro-ph},
       adsurl = {https://ui.adsabs.harvard.edu/abs/2002MNRAS.334..646K}
}

@ARTICLE{white1978,
       author = {{White}, S.~D.~M. and {Rees}, M.~J.},
        title = "{Core condensation in heavy halos: a two-stage theory for galaxy formation and clustering.}",
      journal = {\mnras},
         year = 1978,
        month = may,
       volume = {183},
        pages = {341-358},
          doi = {10.1093/mnras/183.3.341},
       adsurl = {https://ui.adsabs.harvard.edu/abs/1978MNRAS.183..341W}
}

@ARTICLE{fall1980,
       author = {{Fall}, S.~M. and {Efstathiou}, G.},
        title = "{Formation and rotation of disc galaxies with haloes.}",
      journal = {\mnras},
         year = 1980,
        month = oct,
       volume = {193},
        pages = {189-206},
          doi = {10.1093/mnras/193.2.189},
       adsurl = {https://ui.adsabs.harvard.edu/abs/1980MNRAS.193..189F}
}

@ARTICLE{blumenthal1984,
       author = {{Blumenthal}, G.~R. and {Faber}, S.~M. and {Primack}, J.~R. and {Rees}, M.~J.},
        title = "{Formation of galaxies and large-scale structure with cold dark matter.}",
      journal = {\nat},
         year = 1984,
        month = oct,
       volume = {311},
        pages = {517-525},
          doi = {10.1038/311517a0},
       adsurl = {https://ui.adsabs.harvard.edu/abs/1984Natur.311..517B}
}

@ARTICLE{mo1998,
       author = {{Mo}, H.~J. and {Mao}, Shude and {White}, Simon D.~M.},
        title = "{The formation of galactic discs}",
      journal = {\mnras},
         year = 1998,
        month = apr,
       volume = {295},
       number = {2},
        pages = {319-336},
          doi = {10.1046/j.1365-8711.1998.01227.x},
archivePrefix = {arXiv},
       eprint = {astro-ph/9707093},
 primaryClass = {astro-ph},
       adsurl = {https://ui.adsabs.harvard.edu/abs/1998MNRAS.295..319M}
}

@ARTICLE{peebles1969,
       author = {{Peebles}, P.~J.~E.},
        title = "{Origin of the Angular Momentum of Galaxies}",
      journal = {\apj},
         year = 1969,
        month = feb,
       volume = {155},
        pages = {393},
          doi = {10.1086/149876},
       adsurl = {https://ui.adsabs.harvard.edu/abs/1969ApJ...155..393P}
}

@ARTICLE{doroshkevich1970,
       author = {{Doroshkevich}, A.~G.},
        title = "{Spatial structure of perturbations and origin of galactic rotation in fluctuation theory}",
      journal = {Astrophysics},
         year = 1970,
        month = oct,
       volume = {6},
       number = {4},
        pages = {320-330},
          doi = {10.1007/BF01001625},
       adsurl = {https://ui.adsabs.harvard.edu/abs/1970Ap......6..320D}
}

@ARTICLE{white1984,
       author = {{White}, S.~D.~M.},
        title = "{Angular momentum growth in protogalaxies}",
      journal = {\apj},
         year = 1984,
        month = nov,
       volume = {286},
        pages = {38-41},
          doi = {10.1086/162573},
       adsurl = {https://ui.adsabs.harvard.edu/abs/1984ApJ...286...38W}
}

@ARTICLE{weil1998,
       author = {{Weil}, M.~L. and {Eke}, V.~R. and {Efstathiou}, G.},
        title = "{The formation of disc galaxies}",
      journal = {\mnras},
         year = 1998,
        month = nov,
       volume = {300},
       number = {3},
        pages = {773-789},
          doi = {10.1046/j.1365-8711.1998.01931.x},
archivePrefix = {arXiv},
       eprint = {astro-ph/9802311},
 primaryClass = {astro-ph},
       adsurl = {https://ui.adsabs.harvard.edu/abs/1998MNRAS.300..773W}
}

@ARTICLE{sommerlarsen1999,
       author = {{Sommer-Larsen}, Jesper and {Gelato}, Sergio and {Vedel}, Henrik},
        title = "{Formation of Disk Galaxies: Feedback and the Angular Momentum Problem}",
      journal = {\apj},
         year = 1999,
        month = jul,
       volume = {519},
       number = {2},
        pages = {501-512},
          doi = {10.1086/307374},
archivePrefix = {arXiv},
       eprint = {astro-ph/9801094},
 primaryClass = {astro-ph},
       adsurl = {https://ui.adsabs.harvard.edu/abs/1999ApJ...519..501S}
}

@ARTICLE{thacker2000,
       author = {{Thacker}, R.~J. and {Couchman}, H.~M.~P.},
        title = "{Implementing Feedback in Simulations of Galaxy Formation: A Survey of Methods}",
      journal = {\apj},
         year = 2000,
        month = dec,
       volume = {545},
       number = {2},
        pages = {728-752},
          doi = {10.1086/317828},
archivePrefix = {arXiv},
       eprint = {astro-ph/0001276},
 primaryClass = {astro-ph},
       adsurl = {https://ui.adsabs.harvard.edu/abs/2000ApJ...545..728T}
}

@ARTICLE{toomre1972,
       author = {{Toomre}, Alar and {Toomre}, Juri},
        title = "{Galactic Bridges and Tails}",
      journal = {\apj},
         year = 1972,
        month = dec,
       volume = {178},
        pages = {623-666},
          doi = {10.1086/151823},
       adsurl = {https://ui.adsabs.harvard.edu/abs/1972ApJ...178..623T}
}

@INPROCEEDINGS{toomre1977,
       author = {{Toomre}, Alar},
        title = "{Mergers and Some Consequences}",
    booktitle = {Evolution of Galaxies and Stellar Populations},
         year = 1977,
       editor = {{Tinsley}, Beatrice M. and {Larson}, Richard B. Gehret, D. Campbell},
        month = jan,
        pages = {401},
       adsurl = {https://ui.adsabs.harvard.edu/abs/1977egsp.conf..401T}
}

@ARTICLE{quinn1986,
       author = {{Quinn}, P.~J. and {Goodman}, Jeremy},
        title = "{Sinking Satellites of Spiral Systems}",
      journal = {\apj},
         year = 1986,
        month = oct,
       volume = {309},
        pages = {472},
          doi = {10.1086/164619},
       adsurl = {https://ui.adsabs.harvard.edu/abs/1986ApJ...309..472Q}
}

@ARTICLE{quinn1993,
       author = {{Quinn}, P.~J. and {Hernquist}, Lars and {Fullagar}, D.~P.},
        title = "{Heating of Galactic Disks by Mergers}",
      journal = {\apj},
         year = 1993,
        month = jan,
       volume = {403},
        pages = {74},
          doi = {10.1086/172184},
       adsurl = {https://ui.adsabs.harvard.edu/abs/1993ApJ...403...74Q}
}

@ARTICLE{toth1992,
       author = {{Toth}, G. and {Ostriker}, J.~P.},
        title = "{Galactic Disks, Infall, and the Global Value of Omega}",
      journal = {\apj},
         year = 1992,
        month = apr,
       volume = {389},
        pages = {5},
          doi = {10.1086/171185},
       adsurl = {https://ui.adsabs.harvard.edu/abs/1992ApJ...389....5T}
}

@ARTICLE{walker1996,
       author = {{Walker}, Ian R. and {Mihos}, J. Christopher and {Hernquist}, Lars},
        title = "{Quantifying the Fragility of Galactic Disks in Minor Mergers}",
      journal = {\apj},
         year = 1996,
        month = mar,
       volume = {460},
        pages = {121},
          doi = {10.1086/176956},
archivePrefix = {arXiv},
       eprint = {astro-ph/9510052},
 primaryClass = {astro-ph},
       adsurl = {https://ui.adsabs.harvard.edu/abs/1996ApJ...460..121W}
}

@ARTICLE{velazquez1999,
       author = {{Velazquez}, Hector and {White}, Simon D.~M.},
        title = "{Sinking satellites and the heating of galaxy discs}",
      journal = {\mnras},
         year = 1999,
        month = apr,
       volume = {304},
       number = {2},
        pages = {254-270},
          doi = {10.1046/j.1365-8711.1999.02354.x},
archivePrefix = {arXiv},
       eprint = {astro-ph/9809412},
 primaryClass = {astro-ph},
       adsurl = {https://ui.adsabs.harvard.edu/abs/1999MNRAS.304..254V}
}

@ARTICLE{bournaud2007,
       author = {{Bournaud}, F. and {Jog}, C.~J. and {Combes}, F.},
        title = "{Multiple minor mergers: formation of elliptical galaxies and constraints for the growth of spiral disks}",
      journal = {\aap},
         year = 2007,
        month = dec,
       volume = {476},
       number = {3},
        pages = {1179-1190},
          doi = {10.1051/0004-6361:20078010},
archivePrefix = {arXiv},
       eprint = {0709.3439},
 primaryClass = {astro-ph},
       adsurl = {https://ui.adsabs.harvard.edu/abs/2007A&A...476.1179B}
}

@ARTICLE{springel2005,
       author = {{Springel}, Volker and {Hernquist}, Lars},
        title = "{Formation of a Spiral Galaxy in a Major Merger}",
      journal = {\apjl},
         year = 2005,
        month = mar,
       volume = {622},
       number = {1},
        pages = {L9-L12},
          doi = {10.1086/429486},
archivePrefix = {arXiv},
       eprint = {astro-ph/0411379},
 primaryClass = {astro-ph},
       adsurl = {https://ui.adsabs.harvard.edu/abs/2005ApJ...622L...9S}
}

@ARTICLE{athanassoula2016,
       author = {{Athanassoula}, E. and {Rodionov}, S.~A. and {Peschken}, N. and {Lambert}, J.~C.},
        title = "{Forming Disk Galaxies in Wet Major Mergers. I. Three Fiducial Examples}",
      journal = {\apj},
         year = 2016,
        month = apr,
       volume = {821},
       number = {2},
          eid = {90},
        pages = {90},
          doi = {10.3847/0004-637X/821/2/90},
archivePrefix = {arXiv},
       eprint = {1602.03189},
 primaryClass = {astro-ph.GA},
       adsurl = {https://ui.adsabs.harvard.edu/abs/2016ApJ...821...90A}
}

@ARTICLE{kruit1981,
       author = {{van der Kruit}, P.~C. and {Searle}, L.},
        title = "{Surface photometry of edge-on spiral galaxies. I - A model for the three-dimensional distribution of light in galactic disks.}",
      journal = {\aap},
         year = 1981,
        month = feb,
       volume = {95},
        pages = {105-115},
       adsurl = {https://ui.adsabs.harvard.edu/abs/1981A&A....95..105V}
}

@ARTICLE{devau1948,
       author = {{de Vaucouleurs}, Gerard},
        title = "{Recherches sur les Nebuleuses Extragalactiques}",
      journal = {Annales d'Astrophysique},
         year = 1948,
        month = jan,
       volume = {11},
        pages = {247},
       adsurl = {https://ui.adsabs.harvard.edu/abs/1948AnAp...11..247D}
}

@ARTICLE{freeman1970,
       author = {{Freeman}, K.~C.},
        title = "{On the Disks of Spiral and S0 Galaxies}",
      journal = {\apj},
         year = 1970,
        month = jun,
       volume = {160},
        pages = {811},
          doi = {10.1086/150474},
       adsurl = {https://ui.adsabs.harvard.edu/abs/1970ApJ...160..811F}
}

@ARTICLE{fukugita1996,
       author = {{Fukugita}, M. and {Ichikawa}, T. and {Gunn}, J.~E. and {Doi}, M. and {Shimasaku}, K. and {Schneider}, D.~P.},
        title = "{The Sloan Digital Sky Survey Photometric System}",
      journal = {\aj},
         year = 1996,
        month = apr,
       volume = {111},
        pages = {1748},
          doi = {10.1086/117915},
       adsurl = {https://ui.adsabs.harvard.edu/abs/1996AJ....111.1748F}
}

@ARTICLE{smith2002,
       author = {{Smith}, J. Allyn and {Tucker}, Douglas L. and {Kent}, Stephen and {Richmond}, Michael W. and {Fukugita}, Masataka and {Ichikawa}, Takashi and {Ichikawa}, Shin-ichi and {Jorgensen}, Anders M. and {Uomoto}, Alan and {Gunn}, James E. and {Hamabe}, Masaru and {Watanabe}, Masaru and {Tolea}, Alin and {Henden}, Arne and {Annis}, James and {Pier}, Jeffrey R. and {McKay}, Timothy A. and {Brinkmann}, Jon and {Chen}, Bing and {Holtzman}, Jon and {Shimasaku}, Kazuhiro and {York}, Donald G.},
        title = "{The u'g'r'i'z' Standard-Star System}",
      journal = {\aj},
         year = 2002,
        month = apr,
       volume = {123},
       number = {4},
        pages = {2121-2144},
          doi = {10.1086/339311},
archivePrefix = {arXiv},
       eprint = {astro-ph/0201143},
 primaryClass = {astro-ph},
       adsurl = {https://ui.adsabs.harvard.edu/abs/2002AJ....123.2121S}
}

@ARTICLE{ivezic2004,
       author = {{Ivezi{\'c}}, {\v{Z}}. and {Lupton}, R.~H. and {Schlegel}, D. and {Boroski}, B. and {Adelman-McCarthy}, J. and {Yanny}, B. and {Kent}, S. and {Stoughton}, C. and {Finkbeiner}, D. and {Padmanabhan}, N. and {Rockosi}, C.~M. and {Gunn}, J.~E. and {Knapp}, G.~R. and {Strauss}, M.~A. and {Richards}, G.~T. and {Eisenstein}, D. and {Nicinski}, T. and {Kleinman}, S.~J. and {Krzesinski}, J. and {Newman}, P.~R. and {Snedden}, S. and {Thakar}, A.~R. and {Szalay}, A. and {Munn}, J.~A. and {Smith}, J.~A. and {Tucker}, D. and {Lee}, B.~C.},
        title = "{SDSS data management and photometric quality assessment}",
      journal = {Astronomische Nachrichten},
         year = 2004,
        month = oct,
       volume = {325},
       number = {6},
        pages = {583-589},
          doi = {10.1002/asna.200410285},
archivePrefix = {arXiv},
       eprint = {astro-ph/0410195},
 primaryClass = {astro-ph},
       adsurl = {https://ui.adsabs.harvard.edu/abs/2004AN....325..583I}
}

@ARTICLE{york2000,
       author = {{York}, Donald G. and {Adelman}, J. and {Anderson}, John E., Jr. and {Anderson}, Scott F. and {Annis}, James and {Bahcall}, Neta A. and {Bakken}, J.~A. and {Barkhouser}, Robert and {Bastian}, Steven and {Berman}, Eileen and {Boroski}, William N. and {Bracker}, Steve and {Briegel}, Charlie and {Briggs}, John W. and {Brinkmann}, J. and {Brunner}, Robert and {Burles}, Scott and {Carey}, Larry and {Carr}, Michael A. and {Castander}, Francisco J. and {Chen}, Bing and {Colestock}, Patrick L. and {Connolly}, A.~J. and {Crocker}, J.~H. and {Csabai}, Istv{\'a}n and {Czarapata}, Paul C. and {Davis}, John Eric and {Doi}, Mamoru and {Dombeck}, Tom and {Eisenstein}, Daniel and {Ellman}, Nancy and {Elms}, Brian R. and {Evans}, Michael L. and {Fan}, Xiaohui and {Federwitz}, Glenn R. and {Fiscelli}, Larry and {Friedman}, Scott and {Frieman}, Joshua A. and {Fukugita}, Masataka and {Gillespie}, Bruce and {Gunn}, James E. and {Gurbani}, Vijay K. and {de Haas}, Ernst and {Haldeman}, Merle and {Harris}, Frederick H. and {Hayes}, J. and {Heckman}, Timothy M. and {Hennessy}, G.~S. and {Hindsley}, Robert B. and {Holm}, Scott and {Holmgren}, Donald J. and {Huang}, Chi-hao and {Hull}, Charles and {Husby}, Don and {Ichikawa}, Shin-Ichi and {Ichikawa}, Takashi and {Ivezi{\'c}}, {\v{Z}}eljko and {Kent}, Stephen and {Kim}, Rita S.~J. and {Kinney}, E. and {Klaene}, Mark and {Kleinman}, A.~N. and {Kleinman}, S. and {Knapp}, G.~R. and {Korienek}, John and {Kron}, Richard G. and {Kunszt}, Peter Z. and {Lamb}, D.~Q. and {Lee}, B. and {Leger}, R. French and {Limmongkol}, Siriluk and {Lindenmeyer}, Carl and {Long}, Daniel C. and {Loomis}, Craig and {Loveday}, Jon and {Lucinio}, Rich and {Lupton}, Robert H. and {MacKinnon}, Bryan and {Mannery}, Edward J. and {Mantsch}, P.~M. and {Margon}, Bruce and {McGehee}, Peregrine and {McKay}, Timothy A. and {Meiksin}, Avery and {Merelli}, Aronne and {Monet}, David G. and {Munn}, Jeffrey A. and {Narayanan}, Vijay K. and {Nash}, Thomas and {Neilsen}, Eric and {Neswold}, Rich and {Newberg}, Heidi Jo and {Nichol}, R.~C. and {Nicinski}, Tom and {Nonino}, Mario and {Okada}, Norio and {Okamura}, Sadanori and {Ostriker}, Jeremiah P. and {Owen}, Russell and {Pauls}, A. George and {Peoples}, John and {Peterson}, R.~L. and {Petravick}, Donald and {Pier}, Jeffrey R. and {Pope}, Adrian and {Pordes}, Ruth and {Prosapio}, Angela and {Rechenmacher}, Ron and {Quinn}, Thomas R. and {Richards}, Gordon T. and {Richmond}, Michael W. and {Rivetta}, Claudio H. and {Rockosi}, Constance M. and {Ruthmansdorfer}, Kurt and {Sandford}, Dale and {Schlegel}, David J. and {Schneider}, Donald P. and {Sekiguchi}, Maki and {Sergey}, Gary and {Shimasaku}, Kazuhiro and {Siegmund}, Walter A. and {Smee}, Stephen and {Smith}, J. Allyn and {Snedden}, S. and {Stone}, R. and {Stoughton}, Chris and {Strauss}, Michael A. and {Stubbs}, Christopher and {SubbaRao}, Mark and {Szalay}, Alexander S. and {Szapudi}, Istvan and {Szokoly}, Gyula P. and {Thakar}, Anirudda R. and {Tremonti}, Christy and {Tucker}, Douglas L. and {Uomoto}, Alan and {Vanden Berk}, Dan and {Vogeley}, Michael S. and {Waddell}, Patrick and {Wang}, Shu-i. and {Watanabe}, Masaru and {Weinberg}, David H. and {Yanny}, Brian and {Yasuda}, Naoki and {SDSS Collaboration}},
        title = "{The Sloan Digital Sky Survey: Technical Summary}",
      journal = {\aj},
         year = 2000,
        month = sep,
       volume = {120},
       number = {3},
        pages = {1579-1587},
          doi = {10.1086/301513},
archivePrefix = {arXiv},
       eprint = {astro-ph/0006396},
 primaryClass = {astro-ph},
       adsurl = {https://ui.adsabs.harvard.edu/abs/2000AJ....120.1579Y}
}

@ARTICLE{stoughton2002,
       author = {{Stoughton}, Chris and {Lupton}, Robert H. and {Bernardi}, Mariangela and {Blanton}, Michael R. and {Burles}, Scott and {Castander}, Francisco J. and {Connolly}, A.~J. and {Eisenstein}, Daniel J. and {Frieman}, Joshua A. and {Hennessy}, G.~S. and {Hindsley}, Robert B. and {Ivezi{\'c}}, {\v{Z}}eljko and {Kent}, Stephen and {Kunszt}, Peter Z. and {Lee}, Brian C. and {Meiksin}, Avery and {Munn}, Jeffrey A. and {Newberg}, Heidi Jo and {Nichol}, R.~C. and {Nicinski}, Tom and {Pier}, Jeffrey R. and {Richards}, Gordon T. and {Richmond}, Michael W. and {Schlegel}, David J. and {Smith}, J. Allyn and {Strauss}, Michael A. and {SubbaRao}, Mark and {Szalay}, Alexander S. and {Thakar}, Aniruddha R. and {Tucker}, Douglas L. and {Vanden Berk}, Daniel E. and {Yanny}, Brian and {Adelman}, Jennifer K. and {Anderson}, John E., Jr. and {Anderson}, Scott F. and {Annis}, James and {Bahcall}, Neta A. and {Bakken}, J.~A. and {Bartelmann}, Matthias and {Bastian}, Steven and {Bauer}, Amanda and {Berman}, Eileen and {B{\"o}hringer}, Hans and {Boroski}, William N. and {Bracker}, Steve and {Briegel}, Charlie and {Briggs}, John W. and {Brinkmann}, J. and {Brunner}, Robert and {Carey}, Larry and {Carr}, Michael A. and {Chen}, Bing and {Christian}, Damian and {Colestock}, Patrick L. and {Crocker}, J.~H. and {Csabai}, Istv{\'a}n and {Czarapata}, Paul C. and {Dalcanton}, Julianne and {Davidsen}, Arthur F. and {Davis}, John Eric and {Dehnen}, Walter and {Dodelson}, Scott and {Doi}, Mamoru and {Dombeck}, Tom and {Donahue}, Megan and {Ellman}, Nancy and {Elms}, Brian R. and {Evans}, Michael L. and {Eyer}, Laurent and {Fan}, Xiaohui and {Federwitz}, Glenn R. and {Friedman}, Scott and {Fukugita}, Masataka and {Gal}, Roy and {Gillespie}, Bruce and {Glazebrook}, Karl and {Gray}, Jim and {Grebel}, Eva K. and {Greenawalt}, Bruce and {Greene}, Gretchen and {Gunn}, James E. and {de Haas}, Ernst and {Haiman}, Zolt{\'a}n and {Haldeman}, Merle and {Hall}, Patrick B. and {Hamabe}, Masaru and {Hansen}, Brad and {Harris}, Frederick H. and {Harris}, Hugh and {Harvanek}, Michael and {Hawley}, Suzanne L. and {Hayes}, J.~J.~E. and {Heckman}, Timothy M. and {Helmi}, Amina and {Henden}, Arne and {Hogan}, Craig J. and {Hogg}, David W. and {Holmgren}, Donald J. and {Holtzman}, Jon and {Huang}, Chih-Hao and {Hull}, Charles and {Ichikawa}, Shin-Ichi and {Ichikawa}, Takashi and {Johnston}, David E. and {Kauffmann}, Guinevere and {Kim}, Rita S.~J. and {Kimball}, Tim and {Kinney}, E. and {Klaene}, Mark and {Kleinman}, S.~J. and {Klypin}, Anatoly and {Knapp}, G.~R. and {Korienek}, John and {Krolik}, Julian and {Kron}, Richard G. and {Krzesi{\'n}ski}, Jurek and {Lamb}, D.~Q. and {Leger}, R. French and {Limmongkol}, Siriluk and {Lindenmeyer}, Carl and {Long}, Daniel C. and {Loomis}, Craig and {Loveday}, Jon and {MacKinnon}, Bryan and {Mannery}, Edward J. and {Mantsch}, P.~M. and {Margon}, Bruce and {McGehee}, Peregrine and {McKay}, Timothy A. and {McLean}, Brian and {Menou}, Kristen and {Merelli}, Aronne and {Mo}, H.~J. and {Monet}, David G. and {Nakamura}, Osamu and {Narayanan}, Vijay K. and {Nash}, Thomas and {Neilsen}, Eric H., Jr. and {Newman}, Peter R. and {Nitta}, Atsuko and {Odenkirchen}, Michael and {Okada}, Norio and {Okamura}, Sadanori and {Ostriker}, Jeremiah P. and {Owen}, Russell and {Pauls}, A. George and {Peoples}, John and {Peterson}, R.~S. and {Petravick}, Donald and {Pope}, Adrian and {Pordes}, Ruth and {Postman}, Marc and {Prosapio}, Angela and {Quinn}, Thomas R. and {Rechenmacher}, Ron and {Rivetta}, Claudio H. and {Rix}, Hans-Walter and {Rockosi}, Constance M. and {Rosner}, Robert and {Ruthmansdorfer}, Kurt and {Sandford}, Dale and {Schneider}, Donald P. and {Scranton}, Ryan and {Sekiguchi}, Maki and {Sergey}, Gary and {Sheth}, Ravi and {Shimasaku}, Kazuhiro and {Smee}, Stephen and {Snedden}, Stephanie A. and {Stebbins}, Albert and {Stubbs}, Christopher and {Szapudi}, Istv{\'a}n and {Szkody}, Paula and {Szokoly}, Gyula P. and {Tabachnik}, Serge and {Tsvetanov}, Zlatan and {Uomoto}, Alan and {Vogeley}, Michael S. and {Voges}, Wolfgang and {Waddell}, Patrick and {Walterbos}, Ren{\'e} and {Wang}, Shu-i. and {Watanabe}, Masaru and {Weinberg}, David H. and {White}, Richard L. and {White}, Simon D.~M. and {Wilhite}, Brian and {Wolfe}, David and {Yasuda}, Naoki and {York}, Donald G. and {Zehavi}, Idit and {Zheng}, Wei},
        title = "{Sloan Digital Sky Survey: Early Data Release}",
      journal = {\aj},
         year = 2002,
        month = jan,
       volume = {123},
       number = {1},
        pages = {485-548},
          doi = {10.1086/324741},
       adsurl = {https://ui.adsabs.harvard.edu/abs/2002AJ....123..485S}
}

@ARTICLE{gunn1998,
       author = {{Gunn}, J.~E. and {Carr}, M. and {Rockosi}, C. and {Sekiguchi}, M. and {Berry}, K. and {Elms}, B. and {de Haas}, E. and {Ivezi{\'c}}, {\v{Z}} . and {Knapp}, G. and {Lupton}, R. and {Pauls}, G. and {Simcoe}, R. and {Hirsch}, R. and {Sanford}, D. and {Wang}, S. and {York}, D. and {Harris}, F. and {Annis}, J. and {Bartozek}, L. and {Boroski}, W. and {Bakken}, J. and {Haldeman}, M. and {Kent}, S. and {Holm}, S. and {Holmgren}, D. and {Petravick}, D. and {Prosapio}, A. and {Rechenmacher}, R. and {Doi}, M. and {Fukugita}, M. and {Shimasaku}, K. and {Okada}, N. and {Hull}, C. and {Siegmund}, W. and {Mannery}, E. and {Blouke}, M. and {Heidtman}, D. and {Schneider}, D. and {Lucinio}, R. and {Brinkman}, J.},
        title = "{The Sloan Digital Sky Survey Photometric Camera}",
      journal = {\aj},
         year = 1998,
        month = dec,
       volume = {116},
       number = {6},
        pages = {3040-3081},
          doi = {10.1086/300645},
archivePrefix = {arXiv},
       eprint = {astro-ph/9809085},
 primaryClass = {astro-ph},
       adsurl = {https://ui.adsabs.harvard.edu/abs/1998AJ....116.3040G}
}

@ARTICLE{hogg2001,
       author = {{Hogg}, David W. and {Finkbeiner}, Douglas P. and {Schlegel}, David J. and {Gunn}, James E.},
        title = "{A Photometricity and Extinction Monitor at the Apache Point Observatory}",
      journal = {\aj},
         year = 2001,
        month = oct,
       volume = {122},
       number = {4},
        pages = {2129-2138},
          doi = {10.1086/323103},
archivePrefix = {arXiv},
       eprint = {astro-ph/0106511},
 primaryClass = {astro-ph},
       adsurl = {https://ui.adsabs.harvard.edu/abs/2001AJ....122.2129H}
}

@ARTICLE{pier2003,
       author = {{Pier}, Jeffrey R. and {Munn}, Jeffrey A. and {Hindsley}, Robert B. and {Hennessy}, G.~S. and {Kent}, Stephen M. and {Lupton}, Robert H. and {Ivezi{\'c}}, {\v{Z}}eljko},
        title = "{Astrometric Calibration of the Sloan Digital Sky Survey}",
      journal = {\aj},
         year = 2003,
        month = mar,
       volume = {125},
       number = {3},
        pages = {1559-1579},
          doi = {10.1086/346138},
archivePrefix = {arXiv},
       eprint = {astro-ph/0211375},
 primaryClass = {astro-ph},
       adsurl = {https://ui.adsabs.harvard.edu/abs/2003AJ....125.1559P}
}

@ARTICLE{lawrence2007,
       author = {{Lawrence}, A. and {Warren}, S.~J. and {Almaini}, O. and {Edge}, A.~C. and {Hambly}, N.~C. and {Jameson}, R.~F. and {Lucas}, P. and {Casali}, M. and {Adamson}, A. and {Dye}, S. and {Emerson}, J.~P. and {Foucaud}, S. and {Hewett}, P. and {Hirst}, P. and {Hodgkin}, S.~T. and {Irwin}, M.~J. and {Lodieu}, N. and {McMahon}, R.~G. and {Simpson}, C. and {Smail}, I. and {Mortlock}, D. and {Folger}, M.},
        title = "{The UKIRT Infrared Deep Sky Survey (UKIDSS)}",
      journal = {\mnras},
         year = 2007,
        month = aug,
       volume = {379},
       number = {4},
        pages = {1599-1617},
          doi = {10.1111/j.1365-2966.2007.12040.x},
archivePrefix = {arXiv},
       eprint = {astro-ph/0604426},
 primaryClass = {astro-ph},
       adsurl = {https://ui.adsabs.harvard.edu/abs/2007MNRAS.379.1599L}
}

@ARTICLE{casali2007,
       author = {{Casali}, M. and {Adamson}, A. and {Alves de Oliveira}, C. and {Almaini}, O. and {Burch}, K. and {Chuter}, T. and {Elliot}, J. and {Folger}, M. and {Foucaud}, S. and {Hambly}, N. and {Hastie}, M. and {Henry}, D. and {Hirst}, P. and {Irwin}, M. and {Ives}, D. and {Lawrence}, A. and {Laidlaw}, K. and {Lee}, D. and {Lewis}, J. and {Lunney}, D. and {McLay}, S. and {Montgomery}, D. and {Pickup}, A. and {Read}, M. and {Rees}, N. and {Robson}, I. and {Sekiguchi}, K. and {Vick}, A. and {Warren}, S. and {Woodward}, B.},
        title = "{The UKIRT wide-field camera}",
      journal = {\aap},
         year = 2007,
        month = may,
       volume = {467},
       number = {2},
        pages = {777-784},
          doi = {10.1051/0004-6361:20066514},
       adsurl = {https://ui.adsabs.harvard.edu/abs/2007A&A...467..777C}
}

@ARTICLE{hewett2006,
       author = {{Hewett}, P.~C. and {Warren}, S.~J. and {Leggett}, S.~K. and {Hodgkin}, S.~T.},
        title = "{The UKIRT Infrared Deep Sky Survey ZY JHK photometric system: passbands and synthetic colours}",
      journal = {\mnras},
         year = 2006,
        month = apr,
       volume = {367},
       number = {2},
        pages = {454-468},
          doi = {10.1111/j.1365-2966.2005.09969.x},
archivePrefix = {arXiv},
       eprint = {astro-ph/0601592},
 primaryClass = {astro-ph},
       adsurl = {https://ui.adsabs.harvard.edu/abs/2006MNRAS.367..454H}
}

@ARTICLE{hambly2008,
       author = {{Hambly}, N.~C. and {Collins}, R.~S. and {Cross}, N.~J.~G. and {Mann}, R.~G. and {Read}, M.~A. and {Sutorius}, E.~T.~W. and {Bond}, I. and {Bryant}, J. and {Emerson}, J.~P. and {Lawrence}, A. and {Rimoldini}, L. and {Stewart}, J.~M. and {Williams}, P.~M. and {Adamson}, A. and {Hirst}, P. and {Dye}, S. and {Warren}, S.~J.},
        title = "{The WFCAM Science Archive}",
      journal = {\mnras},
         year = 2008,
        month = feb,
       volume = {384},
       number = {2},
        pages = {637-662},
          doi = {10.1111/j.1365-2966.2007.12700.x},
archivePrefix = {arXiv},
       eprint = {0711.3593},
 primaryClass = {astro-ph},
       adsurl = {https://ui.adsabs.harvard.edu/abs/2008MNRAS.384..637H}
}

@BOOK{sersic1968,
       author = {{Sersic}, Jose Luis},
        title = "{Atlas de Galaxias Australes}",
    publisher = {Cordoba, Argentina: Observatorio Astronomico},
         year = 1968,
       adsurl = {https://ui.adsabs.harvard.edu/abs/1968adga.book.....S}
}

@ARTICLE{ciotti1991,
       author = {{Ciotti}, L.},
        title = "{Stellar systems following the R1/m luminosity law.}",
      journal = {\aap},
         year = 1991,
        month = sep,
       volume = {249},
        pages = {99-106},
       adsurl = {https://ui.adsabs.harvard.edu/abs/1991A&A...249...99C}
}

@ARTICLE{caon1993,
       author = {{Caon}, N. and {Capaccioli}, M. and {D'Onofrio}, M.},
        title = "{On the shape of the light profiles of early-type galaxies.}",
      journal = {\mnras},
         year = 1993,
        month = dec,
       volume = {265},
        pages = {1013-1021},
          doi = {10.1093/mnras/265.4.1013},
archivePrefix = {arXiv},
       eprint = {astro-ph/9309013},
 primaryClass = {astro-ph},
       adsurl = {https://ui.adsabs.harvard.edu/abs/1993MNRAS.265.1013C}
}

@book{scott1992,
       author = {{Scott}, David W},
        title = "{Multivariate density estimation : theory, practice, and visualization}",
    publisher = {New York : Wiley},
         year = 1992,
}

@ARTICLE{blanton2005,
       author = {{Blanton}, Michael R. and {Schlegel}, David J. and {Strauss}, Michael A. and {Brinkmann}, J. and {Finkbeiner}, Douglas and {Fukugita}, Masataka and {Gunn}, James E. and {Hogg}, David W. and {Ivezi{\'c}}, {\v{Z}}eljko and {Knapp}, G.~R. and {Lupton}, Robert H. and {Munn}, Jeffrey A. and {Schneider}, Donald P. and {Tegmark}, Max and {Zehavi}, Idit},
        title = "{New York University Value-Added Galaxy Catalog: A Galaxy Catalog Based on New Public Surveys}",
      journal = {\aj},
         year = 2005,
        month = jun,
       volume = {129},
       number = {6},
        pages = {2562-2578},
          doi = {10.1086/429803},
archivePrefix = {arXiv},
       eprint = {astro-ph/0410166},
 primaryClass = {astro-ph},
       adsurl = {https://ui.adsabs.harvard.edu/abs/2005AJ....129.2562B}
}

@ARTICLE{mosenkov2015,
       author = {{Mosenkov}, A.~V. and {Sotnikova}, N. Ya. and {Reshetnikov}, V.~P. and {Bizyaev}, D.~V. and {Kautsch}, S.~J.},
        title = "{Does the stellar disc flattening depend on the galaxy type?}",
      journal = {\mnras},
         year = 2015,
        month = aug,
       volume = {451},
       number = {3},
        pages = {2376-2389},
          doi = {10.1093/mnras/stv1085},
archivePrefix = {arXiv},
       eprint = {1505.03383},
 primaryClass = {astro-ph.GA},
       adsurl = {https://ui.adsabs.harvard.edu/abs/2015MNRAS.451.2376M}
}

@ARTICLE{yoachim2006,
       author = {{Yoachim}, Peter and {Dalcanton}, Julianne J.},
        title = "{Structural Parameters of Thin and Thick Disks in Edge-on Disk Galaxies}",
      journal = {\aj},
         year = 2006,
        month = jan,
       volume = {131},
       number = {1},
        pages = {226-249},
          doi = {10.1086/497970},
archivePrefix = {arXiv},
       eprint = {astro-ph/0508460},
 primaryClass = {astro-ph},
       adsurl = {https://ui.adsabs.harvard.edu/abs/2006AJ....131..226Y}
}

@ARTICLE{comeron2011,
       author = {{Comer{\'o}n}, S{\'e}bastien and {Elmegreen}, Bruce G. and {Knapen}, Johan H. and {Salo}, Heikki and {Laurikainen}, Eija and {Laine}, Jarkko and {Athanassoula}, E. and {Bosma}, Albert and {Sheth}, Kartik and {Regan}, Michael W. and {Hinz}, Joannah L. and {Gil de Paz}, Armando and {Men{\'e}ndez-Delmestre}, Kar{\'\i}n and {Mizusawa}, Trisha and {Mu{\~n}oz-Mateos}, Juan-Carlos and {Seibert}, Mark and {Kim}, Taehyun and {Elmegreen}, Debra M. and {Gadotti}, Dimitri A. and {Ho}, Luis C. and {Holwerda}, Benne W. and {Lappalainen}, Jani and {Schinnerer}, Eva and {Skibba}, Ramin},
        title = "{Thick Disks of Edge-on Galaxies Seen through the Spitzer Survey of Stellar Structure in Galaxies (S$^{4}$G): Lair of Missing Baryons?}",
      journal = {\apj},
         year = 2011,
        month = nov,
       volume = {741},
       number = {1},
          eid = {28},
        pages = {28},
          doi = {10.1088/0004-637X/741/1/28},
archivePrefix = {arXiv},
       eprint = {1108.0037},
 primaryClass = {astro-ph.CO},
       adsurl = {https://ui.adsabs.harvard.edu/abs/2011ApJ...741...28C}
}

@ARTICLE{martinez2019,
       author = {{Mart{\'\i}nez-Lombilla}, C. and {Knapen}, J.~H.},
        title = "{Properties of extragalactic thick discs recovered from ultra-deep Stripe82 imaging}",
      journal = {\aap},
         year = 2019,
        month = sep,
       volume = {629},
          eid = {A12},
        pages = {A12},
          doi = {10.1051/0004-6361/201935464},
archivePrefix = {arXiv},
       eprint = {1907.02995},
 primaryClass = {astro-ph.GA},
       adsurl = {https://ui.adsabs.harvard.edu/abs/2019A&A...629A..12M}
}


 \appendix

\section{Technical details of image decomposition}
\label{app:tech_decomp}

This appendix provides technical details used in \secref{sec:decomp}, including the PSF and the sky level, along with tests of their reliability and impact.

\subsection{PSF}

\begin{figure*}
    \includegraphics[width=\textwidth]{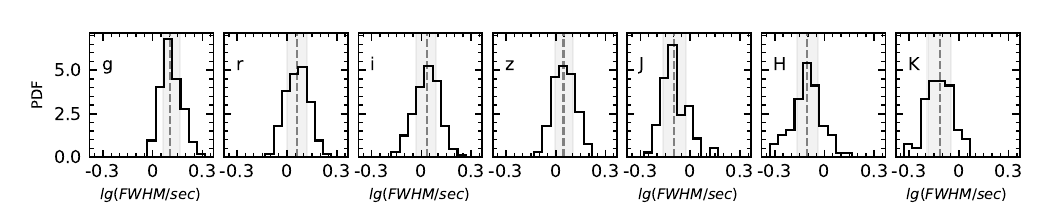}
    \caption{Probability distribution function of PSF FWHM (logarithmic scale) in the SDSS $griz$ and UKIDSS $JHK$ bands. In each panel, the dashed vertical line marks the median value, and the grey shaded region marks the $[0.2, 0.8]$ quantile range.}
    \label{fig:hist_psf}
\end{figure*}

\begin{figure}
    \includegraphics[width=\columnwidth]{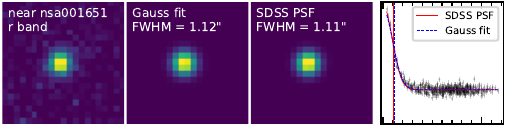}
    \includegraphics[width=\columnwidth]{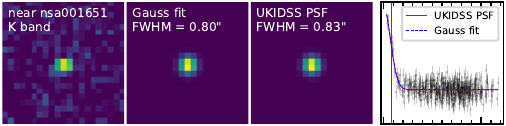}
    \caption{Demonstration of PSF fitting to a point source near a target galaxy. Top and bottom rows correspond to SDSS $r$ band and UKIDSS $K$ band, respectively. Left three panels: the observed image, the fitted model with a Gaussian profile, and the fitted model using the SDSS (or UKIDSS) PSF data. These three images share the same color scale. Rightmost: the radial profiles of these images, where points represent pixels in the observed image, and the red solid and blue dashed curves indicate the SDSS (or UKIDSS) PSF model and the Gaussian fit model, respectively. The vertical lines mark the corresponding half of the PSF FWHM for each model.}
    \label{fig:demo_psf}
\end{figure}

\begin{figure*}
    \includegraphics[width=\textwidth]{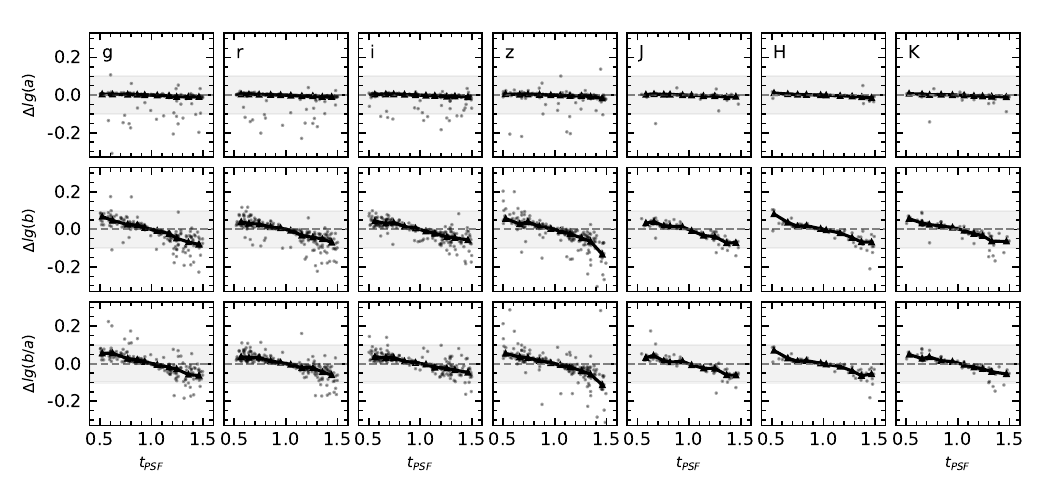}
    \caption{The response of disk shape parameter measurements to the variation in input PSF FWHM. The x-axis $t_{PSF}$ is the ratio of PSF FWHM to that from SDSS/UKIDSS data. The y-axis shows the measurement change with respect to that at $t_{PSF}=1$. In each panel, light grey points correspond to individual measurements, and the heavy black line shows the median value in several x bins, with the median values marked by triangles. The horizontal grey shaded region marks the $\pm$0.1 dex range.}
    \label{fig:comp_rhs_tpsf}
\end{figure*}

In this work, for consistency, a Gaussian profile was adopted to model the PSF in all bands, with the FWHM determined separately for each band. These PSF FWHMs are obtained directly from the SDSS/UKIDSS data services, as one of the outputs of their data processing pipelines. For both surveys, the PSF FWHM represents the average FWHM of well-sampled point sources in the image frame. For the point-source FWHM, SDSS adopts the effective FWHM from a double Gaussian fit, while UKIDSS uses a non-parametric scheme. For further details, we refer the reader to the official documentation of their data pipelines.
\footnote{SDSS: \url{https://www.sdss4.org/dr17/imaging/images}\\
UKIDSS: \url{http://www.ukidss.org/technical/technical.html}}

\figref{fig:hist_psf} shows the distribution of PSF FWHM in the bands used for the image decomposition. The PSF FWHM values are typically around $\sim 1$ arcsec, comparable to the disk minor-axis length $b$ of the superthin galaxies (the second row in \figref{fig:comp_ba_bands}), but they span a narrower range, about 0.3 dex compared with about 0.5 dex in $b$.

To examine the reliability of the SDSS/UKIDSS PSF data, we apply them to point sources near the target superthin galaxies and also fit Gaussian profiles for direct comparison. \figref{fig:demo_psf} shows an example. The profiles match well, with FWHM differences of less than 10\%; the other inspected cases show similar agreement.

\figref{fig:comp_rhs_tpsf} shows tests with variable input PSF FWHM. Each point is a disk-parameter measurement following the same image decomposition procedure as in \secref{sec:decomp}, except that the input PSF FWHM is set to $t_{PSF}\times \mathrm{FWHM}$, where FWHM is the original SDSS/UKIDSS value and $t_{PSF}$ is uniformly sampled from the range $[0.5, 1.5]$. As expected from the relative sizes, changing the PSF has essentially no impact on the disk major-axis length $a$, whereas the minor-axis length $b$ (and hence $b/a$) is affected: a smaller input PSF FWHM yields a larger fitted $b$. Nevertheless, compared with the band-to-band variation of $b$ shown in \figref{fig:comp_ba_bands} (about 0.2 dex between the NIR and optical bands), the PSF-induced change is smaller, reaching about 0.1 dex only when $t_{PSF}$ differs from unity by 0.5. Given the reliability of the SDSS/UKIDSS PSF values demonstrated above, PSF uncertainties do not affect our main conclusions.

\subsection{Sky level}

\begin{figure}
    \includegraphics[width=\columnwidth]{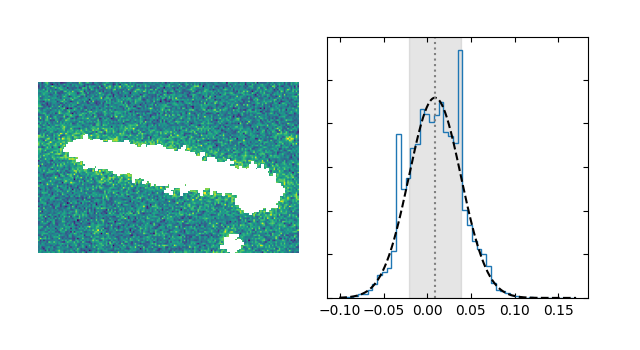}
    \caption{Demonstration of sky level estimation. Left: sky pixels with sources masked; Right: Gaussian profile fit to histogram of the sky pixels.}
    \label{fig:demo_sky_est}
\end{figure}

\begin{figure*}
    \includegraphics[width=\textwidth]{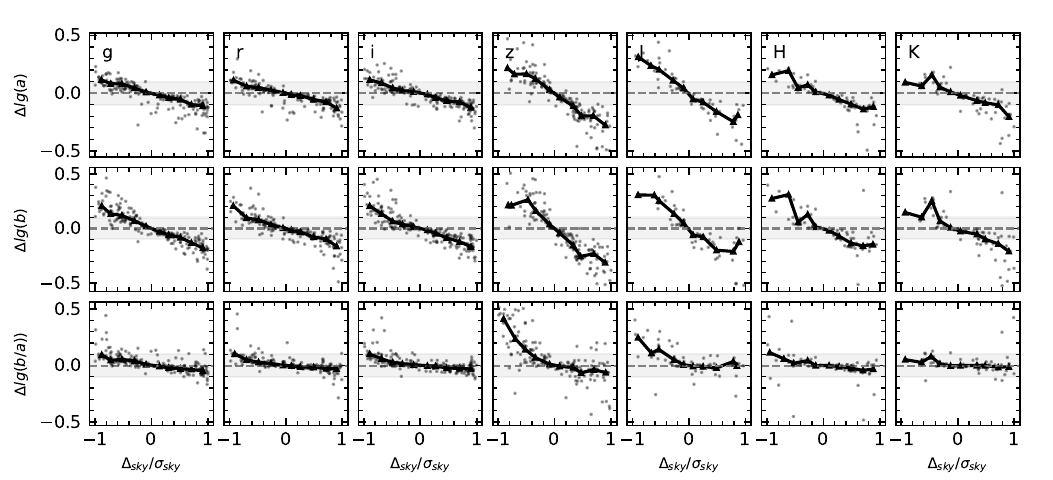}
    \caption{The response of disk shape parameter measurements to the variation in input sky level. The plot is similar to \figref{fig:comp_rhs_tpsf}, except the x-axis is sky-level offset normalized by the sky scatter $\Delta_{sky}/\sigma_{sky}$.}
    \label{fig:comp_rhs_dsky}
\end{figure*}

To estimate the sky level around each galaxy, we fit a Gaussian profile ($\propto \exp(-(I-I_{sky})^2/2 \sigma_{sky}^2)$) to the histogram of background sky pixels, with pixels belonging to foreground sources (identified by \sex) masked out. One example is shown in \figref{fig:demo_sky_est}. Through this procedure, we obtained the sky estimates $I_{sky}$ and $\sigma_{sky}$. For all superthin galaxies involved, we visually inspected their sky estimates to confirm the reliability.

\figref{fig:comp_rhs_dsky} is similar to \figref{fig:comp_rhs_tpsf}, but tests the response of the measurements to variations in the sky level. In these tests, the input sky level is set to $I_{sky} + \Delta_{sky}$, where $I_{sky}$ is the value obtained from the procedure above and $\Delta_{sky}$ is uniformly sampled from the range $[-1, 1] \times \sigma_{sky}$. An underestimated sky level ($\Delta_{sky}<0$) leads to overestimated major- and minor-axis lengths ($a$ and $b$), while an overestimated sky level has the opposite effect. This behaviour is expected: if the subtracted sky level is underestimated, the residual sky background behaves like a very extended source and makes the galaxy appear larger; if the sky level is overestimated, the galaxy profile is artificially truncated. Since the effects on $a$ and $b$ partially cancel, the impact on the ratio $b/a$ is usually less than 0.1 dex. The exception occurs in the $zJ$ bands with $\Delta_{sky} \lesssim -0.5 \sigma_{sky}$, possibly because of their lower signal-to-noise ratios. We therefore conclude that sky-level uncertainties do not significantly affect our main conclusions.

\section{Excluded dust-lane galaxies}
\label{app:dust}

\begin{figure*}
    \includegraphics[width=0.15\textwidth]{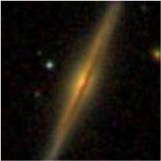}
    \includegraphics[width=0.8\textwidth]{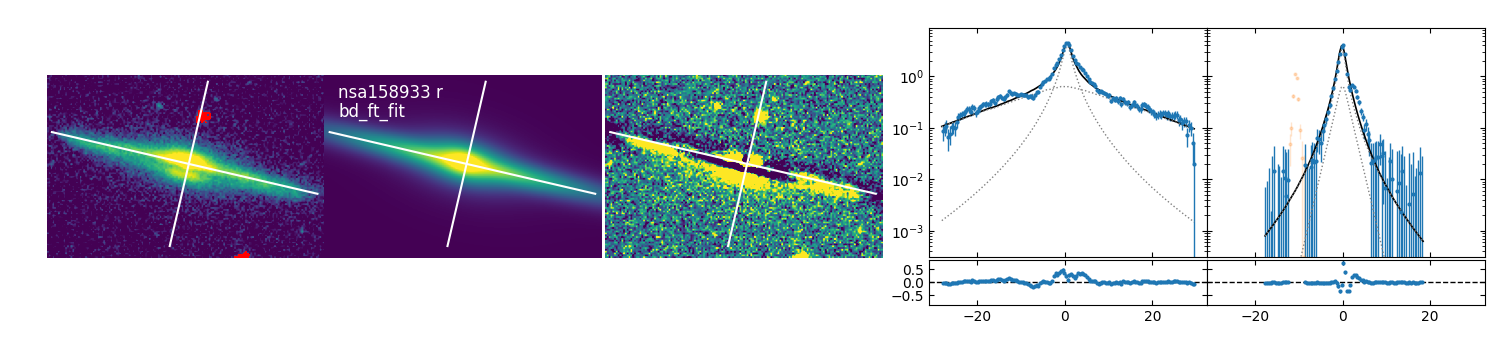}
    \caption{Demonstration of an excluded dust-lane galaxy. Left: SDSS pseudo-color image, similar to the panels in \figref{fig:img_sup_bins}. Right: image modeling of galaxy (observed, model, residual images, and profiles along the major/minor axes), similar to \figref{fig:ellip_galfit}.}
    \label{fig:demo_excl_dust}
\end{figure*}

\begin{figure}
    \includegraphics[width=\columnwidth]{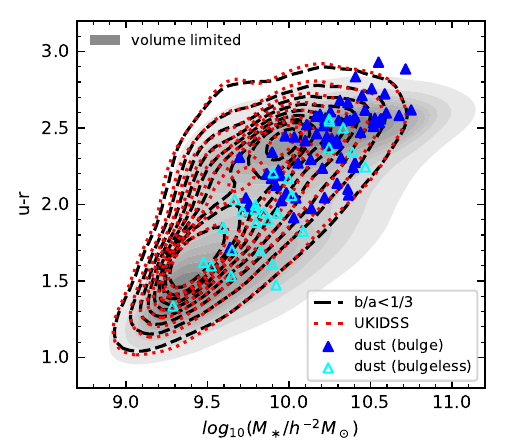}
    \caption{Distribution of excluded dust lane galaxies in the $u-r$ colour versus \lgmstar{} plane, similar to \figref{fig:color_mass}, but with triangles marking the excluded dust-lane galaxies.}
    \label{fig:color_mass_onlydust}
\end{figure}

\begin{figure*}
    \includegraphics[width=\textwidth]{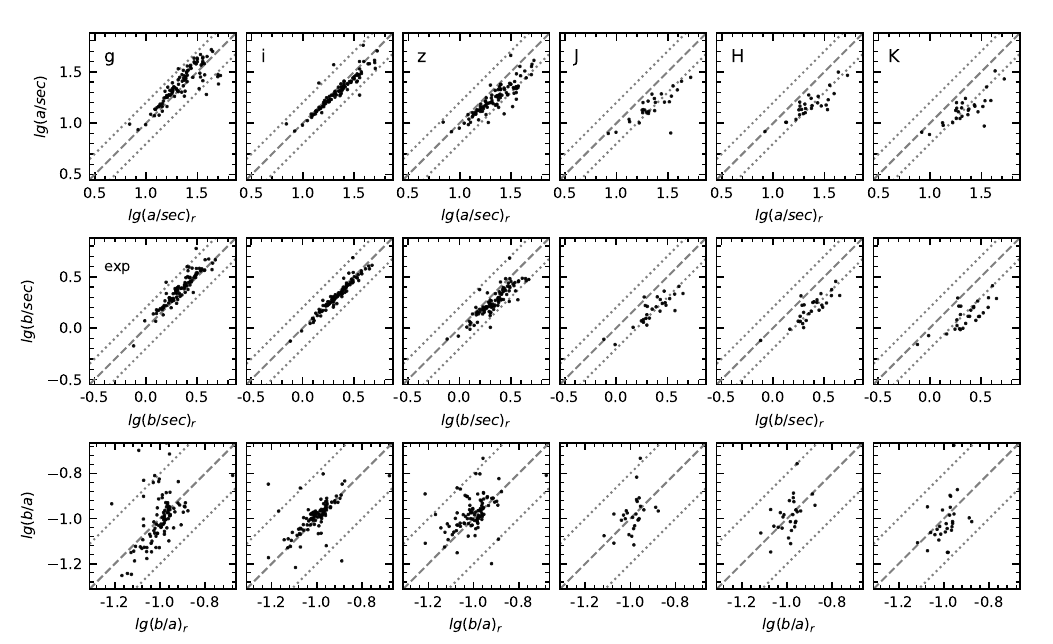}
    \caption{Disk shape parameters of excluded dust-lane galaxies in other bands compared with those in the $r$ band, similar to \figref{fig:comp_ba_bands} and with the same x/y axis ranges.}
	\label{fig:comp_ba_bands_dust}
\end{figure*}

This appendix examines the excluded dust-lane galaxies described in \secref{sec:superthin_sample}.

Our visual exclusions were based on a set of figures, as shown in \figref{fig:demo_excl_dust}. These galaxies exhibit a prominent dust lane in the SDSS pseudo-color image, and residual structure remains after subtraction of the best-fit model. By visually identifying these features, we performed the exclusion described in \secref{sec:superthin_sample}.

\figref{fig:color_mass_onlydust} shows the colour--mass distribution of these excluded dust-lane galaxies. The background distributions are the same as those in \figref{fig:color_mass}. The excluded galaxies lie mostly at the top-right corner, with larger stellar mass and redder colour. The presence of a prominent dust lane directly reddens the observed colour, and tends to overestimate the stellar mass.

\figref{fig:comp_ba_bands_dust} shows disk shape parameters of excluded dust-lane galaxies, similar to \figref{fig:comp_ba_bands}. Although these galaxies have larger sizes along both the major and minor axes, they exhibit comparable size ratios and the same band-dependent trends as the remaining superthin sample discussed in \secref{sec:thin_band}. Including the excluded dust-lane galaxies in the median band-dependence analysis does not significantly change the observed trend.

\section{Construction of the control samples}
\label{app:const_contr}

This appendix describes the construction of the control samples used in \secref{sec:cntr}. The goal is to draw control galaxies whose joint distribution in the relevant control-parameter space matches that of the corresponding superthin sample. Because the superthin sample is small, simple binning in two or more dimensions would be noisy, so we use a kernel-density approach.

For a sample with points $p_i$ in the control-parameter space, we estimate the weighted density as
\begin{equation}
\rho(p)=\sum_i \omega_i K(p; p_i, \sigma),
\label{eq:prob_kde}
\end{equation}
where $K(p; p_i, \sigma)$ is a Gaussian kernel centred on $p_i$ with bandwidth $\sigma$, and $\omega_i$ is the sampling weight assigned to each galaxy. The weights are initially set to unity. The bandwidth is chosen using Scott's rule, as implemented in \texttt{scipy.stats.gaussian\_kde} \citep{scott1992}.

Let $\rho_s(x_k)$ and $\rho_c(x_k)$ be the kernel-density estimates of the parent superthin and control samples on a grid of points $x_k$ in the control-parameter space. We then seek a common target density $\rho(x_k)$ satisfying
\begin{equation}
\begin{aligned}
0 \leq \alpha \rho(x_k) \leq \rho_s(x_k),\\
0 \leq \beta \rho(x_k) \leq \rho_c(x_k),
\end{aligned}
\label{eq:prob_common}
\end{equation}
where $0<\alpha,\beta\leq1$ are the retained fractions of the two samples. We choose $\alpha$ as large as possible in order to retain as many superthin galaxies as possible, while still allowing the control sample to match the same target distribution.

Once the target density is fixed, the sampling weights are chosen so that the weighted density of each parent sample matches the target density. For either parent sample this is solved as the constrained least-squares problem
\begin{equation}
\begin{aligned}
\min \quad & \sum_k \left[t\rho(x_k)-\sum_i \omega_i K(x_k; p_i, \sigma)\right]^2,\\
\mbox{s. t.} \quad & 0 \leq \omega_i \leq 1,
\end{aligned}
\label{eq:weight_resamp}
\end{equation}
where $t$ is fixed to $\alpha$ for the superthin parent sample and to $\beta$ for the control parent sample. We solve this optimization using the convex optimization package CVXOPT\footnote{\url{http://cvxopt.org/index.html}}. The final matched samples are obtained by drawing a random number $q\in[0,1]$ for each galaxy and retaining that galaxy when $q\leq\omega_i$. We tested iterative updates of the bandwidth and weights and found that one iteration is sufficient for the control-sample construction used in this work.

\bsp	
\label{lastpage}
\end{document}